\documentclass{article}
\usepackage[left=1in, right=1in]{geometry}
\usepackage[utf8]{inputenc}
\usepackage{graphicx}
\usepackage{fancyhdr}
\usepackage{gensymb}
\usepackage{mathtools}
\usepackage{ulem}
\graphicspath{{images/}}
\usepackage{amsmath}
\usepackage{bbm}
\usepackage{mathrsfs}
\usepackage{cancel}
\usepackage{amsfonts}
\usepackage{wrapfig,lipsum,booktabs}
\usepackage[american]{babel}
\usepackage{csquotes}
\usepackage[backend=biber]{biblatex}
\title{\vspace{-1.0in} Characterizing Turbulence at a Forest Edge: Comparing Sub-filter Scale Turbulence Models in Simulations of Flow over a Canopy}
\author{Dorianis M. Perez$^{123}$, Jesse M. Canfield $^{3}$, Rodman R. Linn $^{4}$, Kevin Speer $^{12}$}
\date{}

\begin{document}
\maketitle
\vspace{-0.3in}
\begin{center}
\hspace{-0.1in} \textit{$^{1}$Geophysical Fluid Dynamics Institute (GFDI), Florida State University, Tallahassee, FL} \\
\textit{$^{2}$Department of Scientific Computing (DSC), Florida State University, Tallahassee, FL} \\
\textit{$^{3}$X Computational Physics Division, Los Alamos National Laboratory, Los Alamos, NM}
\\
\textit{$^{4}$ Earth and Environmental Science Division, Los Alamos National Laboratory, Los Alamos, NM}
\end{center}

Corresponding author: Dorianis M. Perez dorianisp@lanl.gov

LA-UR-24-20808

\section*{Abstract}
In wildfires, atmospheric turbulence plays a major role in the transfer of turbulent kinetic energy. Understanding how turbulence feeds back into a dynamical system is important, down to the varying small scales of fuel structures (i.e. pine needles, grass). Large eddy simulations (LES) are a common way of numerically representing turbulence. The Smagorinsky model (1963) serves as one of the most studied sub-grid scale representations in LES. In this investigation, the Smagorinsky model was implemented in HIGRAD/FIRETEC, LANL's coupled fire-atmosphere model. The Smagorinsky turbulent kinetic energy (TKE) was compared to FIRETEC's 1.5-order TKE eddy-viscosity subgrid-scale model, known as the Linn turbulence model. This was done in simulations of flow over flat terrain with a homogeneous, cuboidal canopy in the center of the domain. Examinations of the modeled vertical TKE profile and turbulent statistics at the leading edge, and throughout the canopy, show that the Smagorinsky model provides comparable results to that of the original closure model posed in FIRETEC.

\section*{Keywords}
Turbulence, Wildfire, Boundary Layer, Canopy

\section*{Introduction}
Wind speed strongly affects the probability of fire occurrence and increases fire size  (Hernandez et al. 2015). Research investigating the relationship between fire behavior and the motion of the atmosphere around it has become more prevalent since the development of various numerical tools around the turn of the 21st century (Potter, 2012). Turbulence over a canopy has been studied and characterized for decades. Near the surface, between spatial scales on the order of $10^0$ and $10^5$ meters, vegetation exhibits a drag force on the wind field, which decreases wind speeds and creates vertical wind shear as the air well above the canopy continues to move unimpeded (Potter 2012; Holton and Hakim 2012; Pimont et al. 2022). This vertical shear then causes the development of coherent turbulent structures that affect wind flow in and around canopies (Raupach et al. 1996; Pimont et al. 2022). Coherent structures are distinct large-scale fluctuation patterns regularly observed in turbulent flows that are spatially and temporally correlated with themselves and statistically significant with respect to the turbulence energy of the flow (Wilczak 1984, Sullivan et al. 1994).

Numerical modeling of flow through and around a canopy becomes complicated because resolving the spatial and temporal dynamics in fluid motion over the full range of scales -- ranging from those driven by large-scale atmospheric motions down to the scales of vegetation foliage -- would be too computationally expensive (Mueller et al. 2014). A variety of turbulent schemes have been developed and assessed for their ability to reproduce observed turbulence in and around vegetation canopies. These include low-order closure schemes that simulate air flow over and through canopies, yet do not allow for the resolution of canopy turbulence (Dupont and Brunet 2007; Pimont et al. 2009, Li et al. 1990; Green 1992; Liu et al. 1996; Foudhil et al. 2005). When using Large eddy simulation (LES) techniques, the aspects of the flow field that are at scales that are larger than a filter length scale are explicitly resolved and those smaller than the filter length scale are modeled with sub-grid scale parameterization (Pimont et al. 2009). By prescribing a filter width, we resolve turbulent motions that are larger and model those that are smaller (Pope 2000; Mueller et al. 2014). Thus, with a fine enough grid, LES has proven capable of resolving wind gusts in a plant canopy and large coherent eddy structures that may be crucial to wildfire propagation and has been successfully applied over various homogeneous and heterogeneous vegetation canopies under neutral stratification (Pimont et al. 2009; Shaw and Schumann 1992; Kanda and Hino 1994; Su et al., 1998, 2000; Watanabe 2004; Dupont and Brunet 2007, 2008a; Patton et al. 1998; Yang et al. 2006a, 2006b).

FIRETEC (Linn, 1997) is a three-dimensional, physics-based vegetation-drag and combustion model that couples to HIGRAD, a compressible atmospheric hydrodynamics code. Together, HIGRAD and FIRETEC resolve coupled atmosphere-fire behavior with high-fidelity and fire spread on the landscape scale. It represents fire-related phenomenon from sub-grid scale turbulence (Pimont et al., 2009) to generation and propagation of firebrands (Koo et al., 2010, 2012). Studies have shown that HIGRAD/FIRETEC can successfully describe vegetation canopy influences on wind-flow velocity and turbulence (Pimont et al, 2009). A detailed validation study of wind field/canopy interactions in HIGRAD/FIRETEC shows excellent agreement with experiments (Pimont et al., 2009). 

HIGRAD solves the compressible Euler equations in an LES framework for the planetary boundary layer (Reisner et al. 2000a, 2000b). In the context of wildfires, complex turbulent kinetic energy (TKE) transformations occur at various spatial and temporal scales. HIGRAD/FIRETEC employs a turbulence model that uses transport equations to model the turbulence as the sum of three separate turbulence spectra (Linn 1997). The three spectra are modeled with three length scales, each one associated to a sub-filter scale (SFS) TKE conservation equation (Pimont et al. 2009). This model provides a mechanism to transport kinetic energy across disparate scales. However, the three transport equation formulation presents an exceedingly difficult problem in implementing a comprehensive vorticity budget analysis, which is the goal of this work. Thus, it is valuable to examine the potential use of simpler SFS formulations. The choice of SFS stress model in an LES framework plays an important role in the representation of turbulent flow (Mirocha et al. 2010). One of the most commonly used SFS models in LES is the Smagorinsky model (Smagorinsky, 1963), an eddy-viscosity model that relates the residual stress to the filtered rate of strain.

The subject of the present paper is the implementation and evaluation of the Smagorinsky eddy viscosity model in FIRETEC, which comprises the first part of our two-part study. This implementation is motivated by our subsequent study of the vorticity budget equation in HIGRAD/FIRETEC. Using the Smagorinsky model as opposed to the Linn 1.5-order turbulence closure, simplifies the derivation of the vorticity budget equation from the momentum equation, while retaining a sub-grid turbulence closure. The analysis of the vorticity budget equation applied to turbulent flow over a canopy will be discussed in the second part of this study. Thus, the current study serves as a comparison of the Smagorinsky eddy viscosity model to the Linn turbulence model using well-studied parameters and metrics. In the Model description section, HIGRAD/FIRETEC, the Linn turbulence model (Linn, 1997), and the Smagorinsky model are described. The simulation specifics are described in the Methods section. For the Results section, the Smagorinsky model is contrasted against the Linn two-equation, 1.5-order turbulence closure (Linn, 1997). We present results with the Smagorinsky SFS closure for simulations of canopy flow and validate against the Linn closure for the same canopy flow. The Discussion section highlights the model comparison. Concluding remarks and a summary of the findings are provided in the Conclusion.

\subsection*{Model description}
FIRETEC is a three-dimensional, two-phase transport model that conserves mass, momentum, energy, and chemical species written in terrain-following coordinates (Pimont et al. 2009). It includes physics models that predict atmosphere-vegetation interactions, wildfire combustion, heat transfer via convection and radiation, emission of particles like black and organic carbon, and generation and transport of firebrands (Linn et al. 2012; Colman and Linn, 2007; Linn et al. 2005; Josephson et al., 2020; Koo et al. 2012). The equations and corresponding terms are defined in detail in Linn (1997), Linn et al. (2002), and Linn and Cunningham (2005). HIGRAD computes compressible fluid flow in the lower atmospheres, solving Navier-Stokes equations with a large-eddy simulation (LES) approach (Reisner et al. 2000a, 2000b). Governing equations of HIGRAD/FIRETEC are discretized with a finite volume formulation. MPDATA (Smolarkiewicz and Margolin, 1998) is used for advection, the pressure gradient and viscous terms are calculated with central second order differences. The second-order conservative Method of Averages (MOA) approach integrates the model forward in time (Reisner et al., 2000a, 2000b).

The momentum equation in HIGRAD/FIRETEC models unresolved scales through the divergence of a residual stress tensor on the right hand side, 
\begin{equation}
    \frac{\partial}{\partial t} \Big( \rho_g u_i \Big) + \frac{\partial}{\partial x_j} \Big( \rho_g u_i u_j \Big) = - \frac{\partial p}{\partial x_i} + \rho_g g_i - \frac{\partial \tau_{ij}^{r}}{\partial x_j} - \rho_g C_D a_v |u| u_i.
    \label{eq_momentum}
\end{equation}

\noindent where $\rho_g$ is the total gas density, $u_i$ is the gas velocity where $i = 1,2,3$ represents the velocity component in the three Cartesian directions and Einstein summation notation rules are followed. Earth's gravitational acceleration, $ g_i$, is given in the $i$ direction, $\tau_{ij}^{r}$ is the residual stress tensor, $C_D$ is the vegetation drag coefficient, and $a_v$ is the surface area of vegetation per resolved volume.

\subsubsection*{Linn Turbulence Model}
The residual stress tensor, $\tau_{ij}^{r}$, in the Linn turbulence model is the Reynolds stress, $R_{ij}$. The Reynolds stress arises from the Reynolds-Averaged momentum equation posed in Linn (1997), which comes from the advection term, $\rho_g (\textbf{u} \cdot \nabla)\textbf{u}$, being decomposed. Thus, in this formulation, the Reynolds stress, $R_{ij}$, is dependent on the density, $\overline{\rho_g u'_i u'_j}$.
HIGRAD/FIRETEC explicitly resolves turbulent structures from the filter scale up to the domain scale. Smaller turbulent structures are modeled using the  1.5-order Linn turbulence closure scheme (Linn, 1997) that solves SFS turbulent kinetic energy (TKE) conservation equations. Three length scales of turbulence are considered, two of which are associated to SFS TKE conservation equations; equivalent to partitioning the Reynolds stress into three frequency bands, $A$, $B$, and $C$,

\begin{equation}
    R_{ij} = R_{ij,A} + R_{ij,B} + R_{ij,C}.
    \label{eq_Reynolds_Stress}
\end{equation}

\noindent The $R_{ij}$ partitions that are pertinent to modeling wildland fire in FIRETEC are modeled with three length scales, $s_A$, $s_B$, and $s_C$ (Table \ref{variable_table}). Originally, these three length scales were intended to be associated with dominant length scales of vegetation heterogeneity, such as trees or distance between trees, branches or distance between branches, and foliage clusters. However, as the 3D structure and macro-scale heterogeneity of vegetation has become more explicitly modeled, the larger of these scales has taken on a consideration of vegetation resolution and therefore has some dependence on grid resolution. The $A$ scale represents the largest unresolved scale. The $B$ scale is an intermediate scale associated with unresolved structures of vegetation and the smaller scales are binned in $C$. This approach permits these bands of TKE to be modeled with a combination of conservation equations and constitutive relations. The Linn model has successfully described wind field velocity modifications and turbulence in the wildland environments where wildfires persist (Linn and Cunningham, 2005; Cunningham and Linn, 2007; Pimont et al., 2009, Banerjee et al., 2020).

The partitioned Reynolds stress tensors $R_{ij,A}$ and $R_{ij,B}$ respectively are given by,

\begin{equation}
    R_{ij,A} = - \rho_g \nu_t \Bigg( \frac{\partial u_i}{\partial x_j} + \frac{\partial u_j}{\partial x_i} \Bigg) + \frac{2}{3}\delta_{ij}\Bigg(\rho_g \nu_t \frac{\partial u_k}{\partial x_k} + \rho_g K_A \Bigg),
    \label{eq_RijA}
\end{equation}

\begin{equation}
    R_{ij,B} = - \rho_g \nu_t \Bigg( \frac{\partial u_i}{\partial x_j} + \frac{\partial u_j}{\partial x_i} \Bigg) + \frac{2}{3}\delta_{ij}\Bigg(\rho_g \nu_t \frac{\partial u_k}{\partial x_k} + \rho_g K_B \Bigg),
    \label{eq_RijB}
\end{equation}

\noindent where $\nu_t$ is the turbulent viscosity and $\delta_{ij}$ is the Kronecker delta. The eddy viscosity is,

\begin{equation}
    \nu_t = 0.09 \big(s_{A} K_{A}^{1/2} + s_{B} K_{B}^{1/2} +s_{C} K_{C}^{1/2} \big).
    \label{eq_Linn_Eddy_Viscosity}
\end{equation}

\noindent Similar relations to equations (\ref{eq_RijA}) and (\ref{eq_RijB}) could be constructed for $R_{ij,C}$. However, a self-similar approximation for $C$ scales relative to $B$ scales, $R_{ij,C} \propto R_{ij,B}$, has proven sufficient. The transport equations for $K_A$ and $K_B$ are derived by taking moments of velocity with the momentum equation and modeling higher order terms. Those details are captured elsewhere (Linn, 1997). The resulting two-equation turbulence model follows,

\begin{equation}
\begin{split}
    \frac{\partial}{\partial t} ( \rho_g K_A ) + \frac{\partial}{\partial x_i} ( \rho_g K_A u_i ) 
    &= -R_{il,A} \frac{\partial u_i}{\partial x_l} + \frac{2}{3} C_{DR} \frac{\partial}{\partial x_l} \Big( s_A \frac{R_{lk}}{K^{1/2}} \frac{\partial K_A}{\partial x_k} \Big) \\
    &- \frac{K^{1/2}}{s_A} K_A - \rho_g C_D a_{v} K^{1/2} K_A,
\end{split}
\label{eq_Turbulence_A}
\end{equation}

\noindent and,

\begin{equation}
\begin{split}
    \frac{\partial}{\partial t} ( \rho_g K_B ) + \frac{\partial}{\partial x_i} ( \rho_g K_B u_i ) &= -R_{il,B} \frac{\partial u_i}{\partial x_l} + \frac{2}{3} C_{DR} \frac{\partial}{\partial x_l} \Big( s_B \frac{R_{lk}}{K^{1/2}} \frac{\partial K_B}{\partial x_k} \Big) \\
    &- \frac{K^{1/2}}{s_B} K_B - \rho_g C_D a_{v,C} K^{1/2} K_B + \rho_g C_D a_{v} (K^{1/2} K_A + |u|^3 ).
\end{split}
\label{eq_Turbulence_B}
\end{equation}

\noindent In equations (\ref{eq_Turbulence_A}) and (\ref{eq_Turbulence_B}), $C_{DR}$ is a constant, $s_A$ is the characteristic length of the $A$ scales, $s_B$ is the characteristic length of the $B$ scales, $a_{v}$ is the surface area to resolved volume ratio of the fuels, and the total SFS TKE is given by $K = K_A + K_B + K_C$.

\begin{wraptable}{L}{0.5\textwidth}
    \begin{tabular}{c c c}
       Terms & Designation & Value\\
    \hline
       $C_D$  & Drag coefficient & 1.0\\
       $C_{DR}$  & Coefficient for diffusion transport & 2.0\\
       $s_A$  & Largest unresolved scale & 2.0\\
       $s_B$  & Scale of vegetation wake structures & 0.25 \\
       $s_C$  & Scale of fine vegetation & 0.1\\
       $K_A$  & Unresolved Linn TKE at $s_A$ scale & - \\
       $K_B$  & Unresolved Linn TKE at $s_B$ scale & - \\
       $K_C$  & Unresolved Linn TKE at $s_C$ scale & - \\
       $K_S$  & Unresolved Smagorinsky TKE & - \\
       $K_{S,tot}$ & Total Smagorinsky TKE  & - \\
       $K_{L,tot}$ & Total Linn TKE  & - \\
    \hline
    \end{tabular}
    \caption{Constants and variables defined and referenced throughout the Model Description Section.}
    \label{variable_table}
\end{wraptable}

The left hand sides of equations (\ref{eq_Turbulence_A}) and (\ref{eq_Turbulence_B}) describe advection transport of the two turbulent energies. The first terms on the right hand sides represent shear production at those scales, the second terms are the modeled turbulent diffusion transport terms. The third right hand side terms model dissipation to heat for those scales. The fourth term in (\ref{eq_Turbulence_A}) is a sink from vegetation drag at the $A$ scale. The fourth term in (\ref{eq_Turbulence_B}) is a drag sink from the $B$ scale to smaller scales.  In (\ref{eq_Turbulence_B}), the fifth term is the corresponding source term due to the $A$ scale drag sink term. 

As described above, $C$ scales are assumed to be proportional to $B$ scales with $K_C = 0.2K_B$ (Linn, 1997; Linn and Cunningham 2005; Pimont et al. 2009). For the purpose of this investigation, because $A$ scales represent the largest unresolved length scale, the quantities associated with $K_A$ will be used for comparison to the Smagorinsky TKE. The values of these scales are found in Table \ref{variable_table}.

HIGRAD/FIRETEC with the Linn closure has been validated and investigated using various configurations of observational data, from the experimental fires over homogeneous grasslands in Australia (Cheney et al. 1998; Linn and Cunningham 2005) to crown fires of the International Crown Fire Modeling Experiment (ICFME) (Linn et al. 2005b). Pimont et al. (2009) validated HIGRAD/FIRETEC's ability to accurately reproduce essential features of turbulent flow over forests including fuel-breaks. They also performed sensitivity studies to show that the model is not sensitive to uncertain parameters like the vegetation drag coefficient.

\subsubsection*{Smagorinsky SFS Model}

In this study, the Smagorinsky SFS model was implemented in HIGRAD/FIRETEC in place of the Linn model.  In this LES model, the rate of transfer of energy acts from the filtered motions directly to the residual motions through the residual stress tensor, $\tau_{ij}^{r}$. To close the filtered momentum equation, a model for $\tau_{ij}^{r}$ is needed. In the Smagorinsky framework, an eddy viscosity model is used to relate the residual stress to the filtered rate of strain, 

\begin{equation}
    \tau_{ij}^{r} = -2 \rho_g \nu_{r} \overline{S}_{ij}
    \label{eq_Viscosity_Model}
\end{equation}
where the coefficient of proportionality $\nu_{r}$ is the eddy viscosity of the residual motions. It is modeled by, 

\begin{equation}
    \nu_r = \ell_{S}^2 \overline{S} = (C_S \Delta)^2 \overline{S} ,
    \label{eq_Smagorinsky}
\end{equation}

\noindent where $\overline{S}$ is the characteristic filtered rate of strain and is defined by the magnitude of twice the strain rate tensor, $\overline{S} \equiv (2 \bar{S}_{ij}  \bar{S}_{ij})^{\frac{1}{2}}$. In equation (\ref{eq_Smagorinsky}), $\ell_{S}$ is the Smagorinsky length scale which is proportional to the filter width, $\Delta = \sqrt{\Delta x^2 + \Delta y^2 + \Delta z^2}$, through the Smagorinsky coefficient, $C_{S} = 0.17$ (Pope, 2000). $\overline{S}_{ij}$ is the symmetric filtered strain rate tensor ,
\begin{equation}
    \overline{S}_{ij} = \frac{1}{2} \Bigg( \frac{\partial u_i}{\partial x_j} + \frac{\partial u_j}{\partial x_i} \Bigg) - \frac{1}{3} \delta_{ij} \frac{\partial u_{k}}{\partial x_{k}}
\end{equation}
where the second term arises for compressible flow (Mihalas and Weibel-Mihalas 1999). The numerical experiment methods are described next.

\section*{Methods}

\subsection*{Numerical Details}

Pimont et al. (2009) validated the wind dynamics in HIGRAD/FIRETEC over vegetated canopies with a neutral thermal stratification and without considering fire propagation or interaction between winds and strong buoyant sources. The model was validated over a continuous forest canopy using the in-situ measurements of Shaw et al. (1988) and a discontinuous forest using the wind-tunnel measurements of Raupach et al. (1987). Pimont et al. (2006) addressed issues of using Dirichlet boundary conditions for simulating flows in canopies with physics models, as they generate wind velocity and TKE fields that are not physically realistic. As a result, Pimont et al. (2009) used periodic boundary conditions on lateral boundaries in the crosswind direction. A Rayleigh damping layer was used at the upper boundary as well as at lateral boundaries in the streamwise direction to absorb propagating wave disturbances and to eliminate wave reflection at the boundaries. Pimont et al. (2006) pointed out that non-periodic cross-stream boundary conditions for wind computations could provide unrealistic wind profiles. Pre-computed wind fields were used as upstream and downstream boundary conditions on the $x-$ and $y-$inlets, as suggested by Canfield et al. (2005) and Pimont et al. (2009). The pre-computed or precursor data were wind simulations without the tree canopy present that had periodic boundaries in the horizontal directions over a homogeneous layer of grass vegetation in the bottom grid cells. This setup permits the wind field to spin-up a turbulent boundary layer over time and achieve a shear profile with quasi-equilibrium turbulence due to drag from the grass. A snapshot of the thermodynamic state and velocity field were saved once quasi-equilibrium is satisfied. Then, the horizontal domain boundaries were saved at high frequency (10 hz). The snapshot and boundary files were then used as initial and boundary conditions in the canopy runs. A Rayleigh damping layer was used in a 165 m thick layer along the upper boundary. 

\begin{figure}[h!]
         \centering            
         \includegraphics[width=0.75\textwidth]{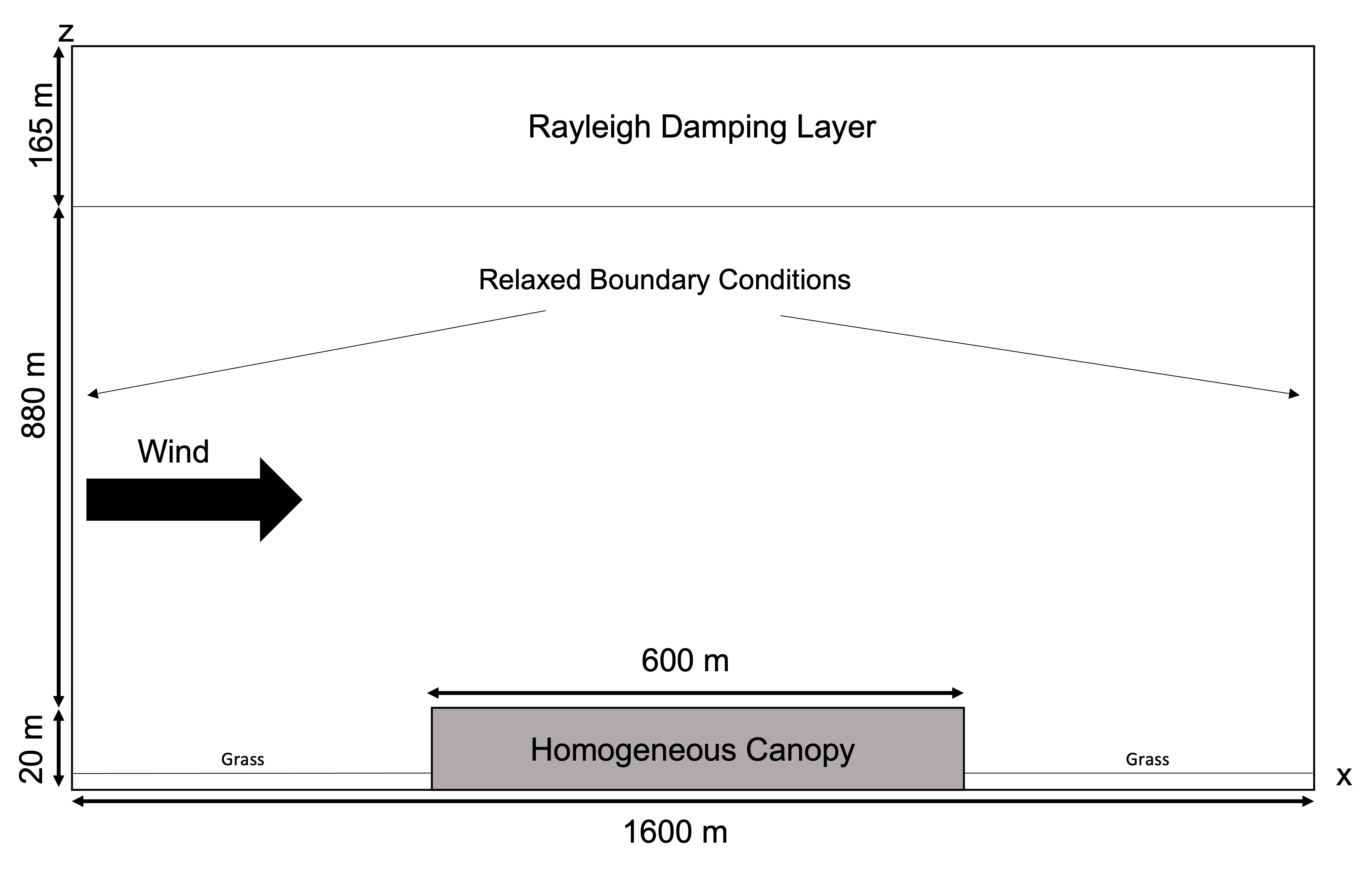}
         \caption{Schematic representation of the computational domain used for the homogeneous canopy simulations.}
         \label{schematic}
\end{figure}

The current computational domain was chosen so that the boundaries have a negligible effect on the dynamics of the flow around the canopy. A homogeneous canopy in the center of the domain with surface vegetation representing a tree canopy with grass all around was used. It should be recognized that this choice of a homogenized canopy could have implications for the similarity of results between Smagorinsky and the Linn Model. The domain size is $1600 \times 1600 \times 1065m$ with a horizontal spacing of $2m$ and vertical mesh stretching starting from $1.5m$ near the ground to $40m$ at the top. The canopy height was set at $20m$ and is $600 \times 600m$ horizontally (Figure \ref{schematic}). The chosen domain size was selected to be large enough to include the largest eddies over the area of interest and the size of the largest eddy is determined by defining an integral length scale that is a measure of the largest correlation distance between two points in the turbulent flow. A uniform vegetation load consistent with dense grass, bulk density of $1.1 \frac{kg}{m^3}$, resides in the bottom $0.27 m$ of the domain. A uniform homogeneous canopy is simulated in the center of the domain (Figure \ref{schematic}). The bulk density of the canopy is $0.0974 \frac{kg}{m^3}$ and is $h=20 m$ in height. The surface area-to-volume ratio is directly proportional to the bulk density by way of the ratio of the cell height to the depth of the fuel. Three additional simulations were performed to confirm the convergence of the Smagorinsky implementation in FIRETEC. This convergence study was performed with grass at grid resolutions of $2m$, $4m$, and $8m$, horizontal periodic boundary conditions, and a Rayleigh damping layer at the upper boundary. 

Each simulation was run on 1600 processors for approximately 100 hours of wall-clock time. No fire propagation was considered. Wind velocity and turbulence statistics profiles were computed from time-averaging (denoted by $\langle \rangle_{t}$) between 1500s and 3000s, at which point the simulations had reached a quasi steady-state, with a 50-second interval between realizations for the averaging procedures. Resolved wind velocity components $u_i$ can be decomposed into $u_i = \langle u'_i \rangle_t + u'_i$. The mean total TKE ($K_{S/L,tot}$) profiles were computed as 
\begin{equation}
  \begin{split}
    K_{L, tot} &= 0.5 \langle u'_i u'_i \rangle_t + \langle K_{A} \rangle_t ,\\
    K_{S, tot} &= 0.5 \langle u'_i u'_i \rangle_t + \langle K_{S} \rangle_t ,
    \label{eq_Resolved_TKE}
  \end{split}
\end{equation}

\noindent where $0.5 \langle u'_i u'_i \rangle_t$ represents the time-averaged resolved Linn or Smagorinsky TKE, respectively, $\langle K_A \rangle_t$ is the time-averaged unresolved Linn TKE at the $A$ scale, and $\langle K_S \rangle_t$ is the time-averaged unresolved Smagorinsky TKE. $K_S$ is solved for from the Prandtl and Kolmogorov formulation of the turbulent diffusion coefficient and assumes that sub-grid scale (SGS) turbulent diffusion is isotropic (Pope 2000)

\begin{equation}
    \nu_r = 0.09\rho_g \Delta K_{S}^{1/2} .
    \label{eq_Smagorinsky_Eddy_Viscosity}
\end{equation}

\noindent Equation (\ref{eq_Smagorinsky_Eddy_Viscosity}) is solved for $K_{S}$ to yield 
\begin{equation}
    K_S = \Big(\frac{\nu_r}{0.09 \rho_g \Delta} \Big)^{2} =  \Big(\frac{C_S^2 \Delta \overline{S}}{0.09 \rho_g} \Big)^{2}
\end{equation}

\noindent for comparison with $K_A$ from the Linn turbulence model. Equations (\ref{eq_Linn_Eddy_Viscosity}) and (\ref{eq_Smagorinsky_Eddy_Viscosity}) are equivalent for their respective length scales and TKEs. The Prandtl/Kolmogorov formulation is extensively used for calculating turbulent viscosity. We solve for $K_S$ from this formulation because it provides an additional, independent equation for calculating the TKE and is consistent with other implementations of the Smagorinsky model (Horiuti and Tamaki 2013; Taghinia 2015).

\section*{Results}
    \begin{figure}[h!]
        \centering        
        \includegraphics[width=11cm]{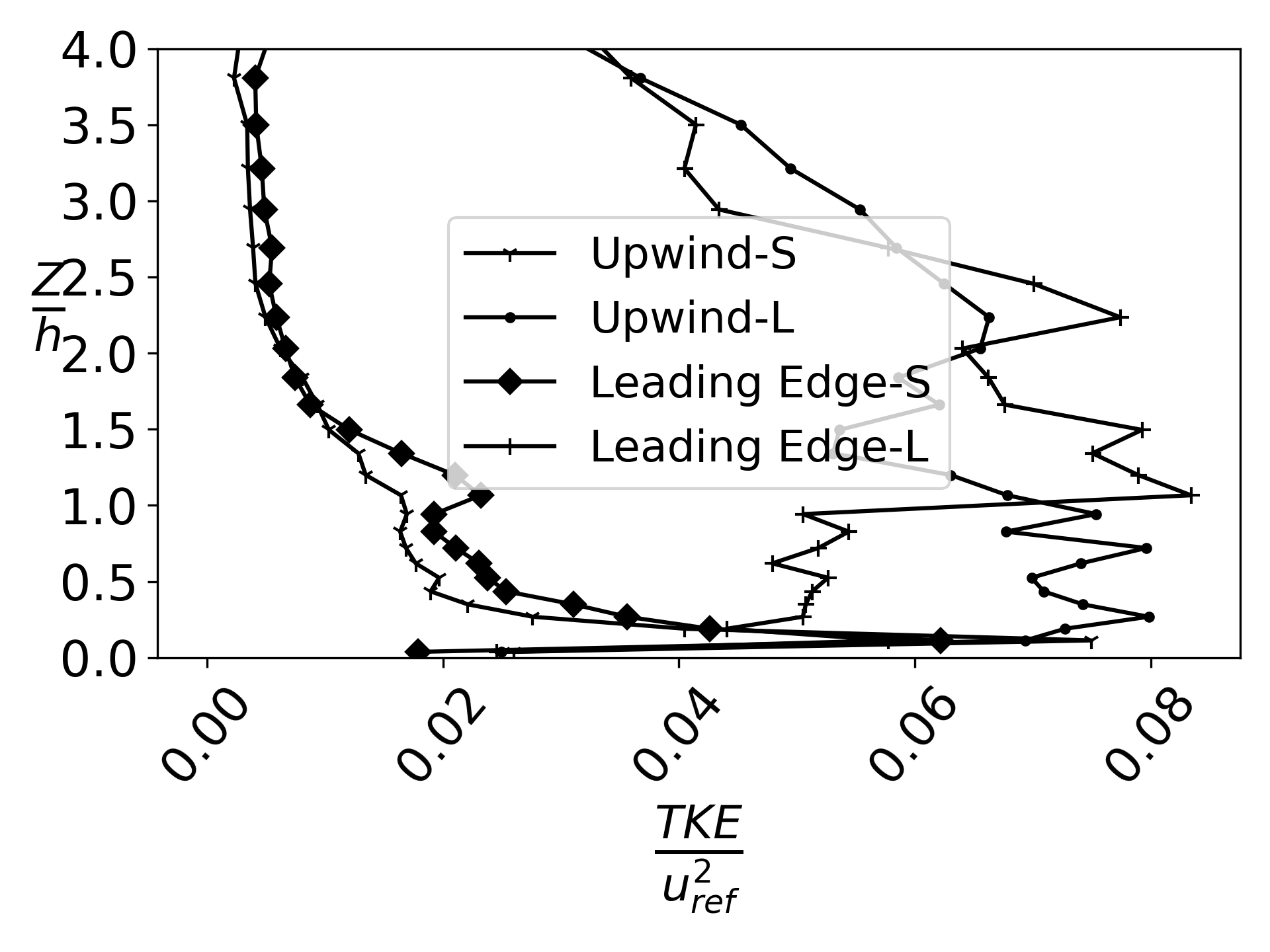}  \includegraphics[width=11cm]{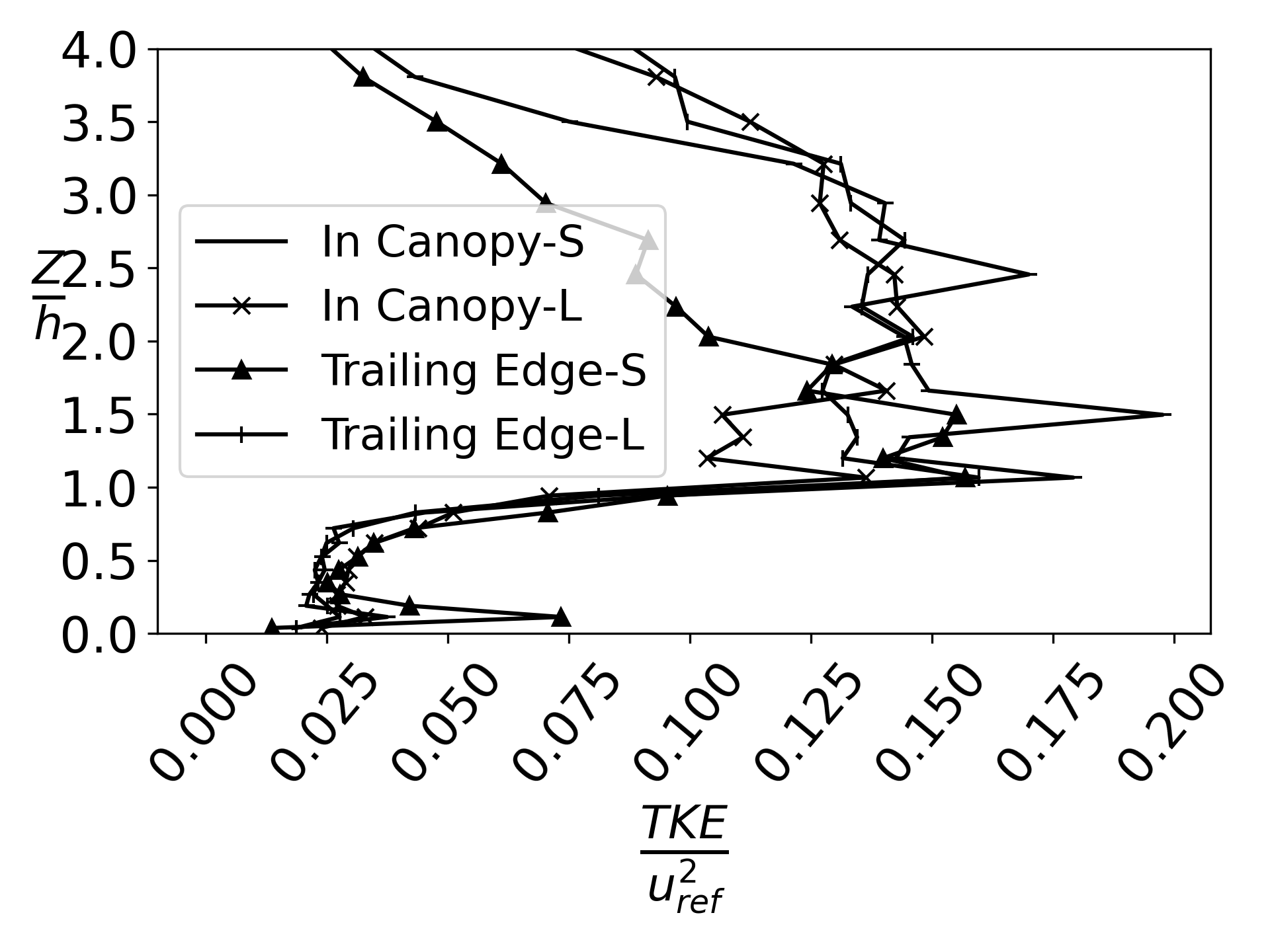}
        \caption{Resolved TKE = $0.5 \langle \tilde{u}'_i \tilde{u}'_i \rangle_t$ vertical profiles, normalized by $u_{ref}^2=U_{in}^2$ upwind of the canopy $(150, 400)$ and at the leading edge $(250,400)$ (top); and in the canopy $(400,400)$ and at the trailing edge $(550,400)$ (bottom) for the Linn model (-L in the legend) and Smagorinsky turbulence models (-S in the legend). }
        \label{verticalTKE}
    \end{figure}

\begin{figure}[h!]
        \centering
     \includegraphics[width=12cm]{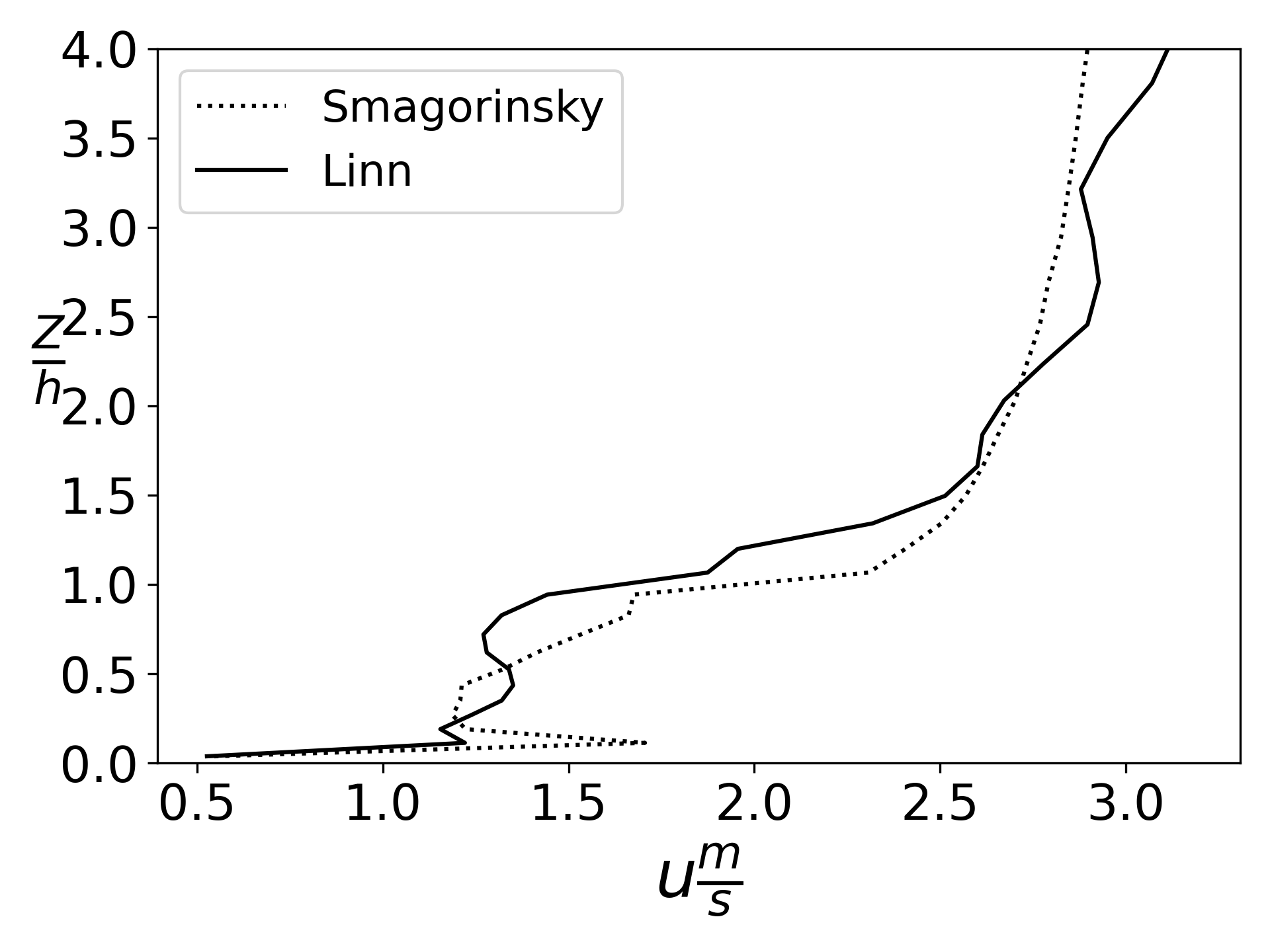}
        \caption{Mean wind profiles at the leading edge of the canopy show the decrease in streamwise velocity due to canopy drag nearing zero in the canopy for both simulations. }
        \label{canopydrag}
\end{figure}

Flow characteristics around the canopy, with particular interest near the leading edge, were analyzed and results from the Smagorinsky model were compared to the Linn turbulence model. The time-averaged vertical profiles of the resolved TKE are compared for both model implementations in Figure \ref{verticalTKE} at various locations in the domain: upwind at the $(x,y)$ point (in meters) $(300,800)$; at the leading edge of the canopy $(500,800)$; in the canopy $(800,800)$; and at the trailing edge of the canopy $(1100,800)$. The TKE values are normalized by the square of the characteristic wind speed, $u_{ref}^2$. Vertical profiles of the TKE upwind and at the leading edge of the canopy contain more turbulence in the Linn model than in the Smagorinsky model. There is a sharp peak near the surface at both of these locations because of the presence of a dense layer of grass that imposes a large drag on the flow. Upwind and at the leading edge, the turbulence tapers to its free-stream value at a height of about $2h$ with the Smagorinsky model. The Linn model produces turbulence profiles that taper to the free-stream values much higher above the canopy. Upstream of the canopy, the differences between the two models in TKE vertical profiles is carried over from the precursor runs, where the Linn model generated more turbulence than the Smagorinsky model. This is observed in the wind fields upstream of the canopy in Figures \ref{paraviewU} and \ref{paraviewW} below. The discrepancy has two potential sources. The Smagorinsky model does not produce backscatter from the unresolved to the resolved scales (Pope, 2000), whereas the Linn model permits backscatter through sources in the momentum equation. The second reason that more kinetic energy is present with the Linn model is that $K_A$ and $K_B$ are persistent variables. That is, unresolved turbulence is transported with partial differential equations that have source and sink terms. In this way, turbulence that is generated by some feature of the flow is carried downstream, interacting with the resolved and unresolved flow field, collecting and dissipating energy along the way. Compare this to the Smagorinsky model, which is an instantaneous calculation that is spatially local. In other words, the Linn model demonstrates hysteresis, whereas the Smagorinsky model does not.

In the canopy and at the trailing edge, the effects of drag from the vegetation are evident within the canopy, where a sharp peak is observed just above the grass layer near the ground. This is caused by the overlapping grass and canopy vegetation in the bottom layer of computational cells. Just above the grass, the turbulence for both models drops to a minimum. Then, above the canopy the turbulence increases through the turbulent boundary layer that has developed downwind from the leading edge. The Smagorinsky model shows the largest values of normalized TKE where sharp vertical velocity gradients exist. These regions standout at the top of the grass, $z/h \approx 0.1$, the top of the canopy, $z/h=1$, and  the top of the turbulent boundary layer, $1.5 < z/h < 2.5$. We expect this behavior from the Smagorinsky model because the deviatoric stress is the sole turbulence production mechanism for that model. At these same locations, the Linn model shows comparable results to Smagorinsky, but the peak values are less dramatic because it is a more dynamic model, exchanging energy between the source, redistribution, and dissipation terms. As pointed out above, the Linn model has a deeper layer of TKE above the canopy than the Smagorinsky model.

The velocity profiles depicted in Figure \ref{canopydrag}, approximate the well-known surface layer logarithmic profile above the canopy. Within the canopy, the vertical profile shows an inflection point near the canopy top that is typical of canopy mean velocity profiles due to strong wind shear in this region. The shear is driven by the drag forces imposed by the canopy on the flow, decreasing the momentum within the canopy. The wind profiles are similar between the two models throughout the domain except for the leading edge profiles, where there are noticeable differences. Above the canopy, the Smagorinsky velocity profile is a smooth logarithmic curve since the turbulent boundary layer above the grass did not extend above the top of the canopy at the leading edge. The Linn model also produces a velocity profile that increases with height above the canopy. However, it contains undulations that deviate from a logarithmic profile. This is due to resolved eddies that have formed in the deeper turbulent boundary layer that developed over the grass.

The Linn model has more resolved turbulence up- and down-stream of the canopy, which can be seen in the contours of stream-wise velocity at $z=h$ in Figure \ref{paraviewU}. Inside the canopy, which has vegetation loading, drag plays a large role in damping the momentum and thus the TKE inside of the canopy (Figure \ref{canopydrag}). Referring back to Figure \ref{verticalTKE}, the, ``In Canopy," and, ``Trailing Edge," profiles are similar in magnitude and shape below $h$. Above $h$, the profiles for both models diverge though. This is due to the deeper turbulent boundary layer in the Linn model.

\begin{figure}[h!]
         \centering
         \includegraphics[width=14cm]{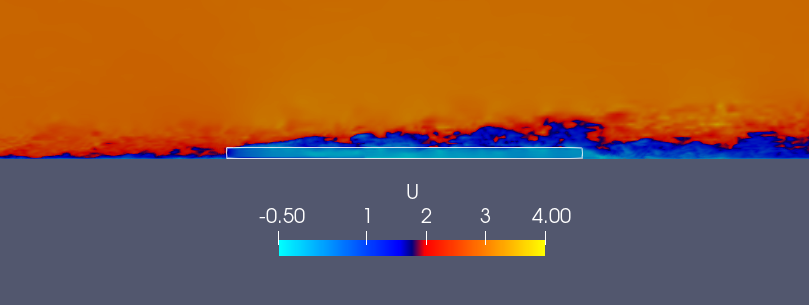}
         \includegraphics[width=14cm]{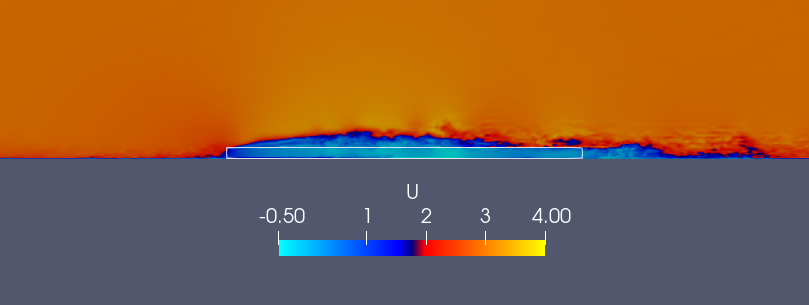}
         \caption{Instantaneous streamwise velocity in a streamwise cross-section through the canopy from the Linn turbulence scheme (top) and Smagorinsky (bottom) at 2000 seconds. The white outline corresponds to the canopy, $600 m$ in length and $20 m$ in height and U is in $\frac{m}{s}$.}
         \label{paraviewU}
\end{figure}

    \begin{figure}[h!]
        \centering
        \includegraphics[width=14cm]{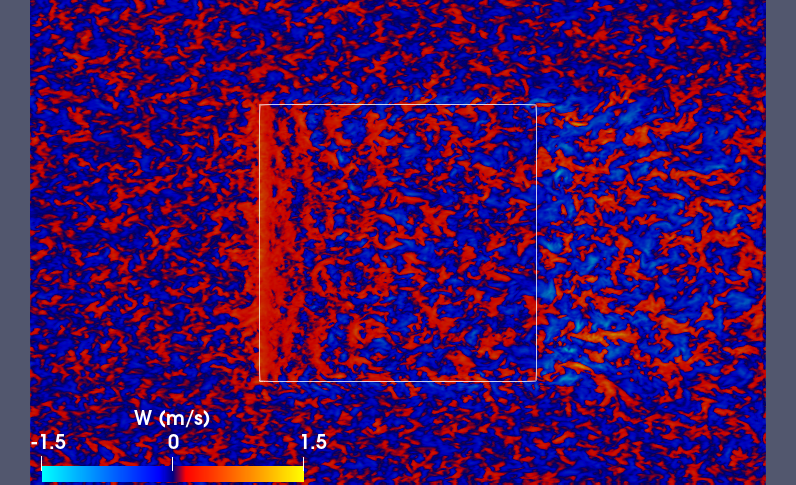}
        \includegraphics[width=14cm]{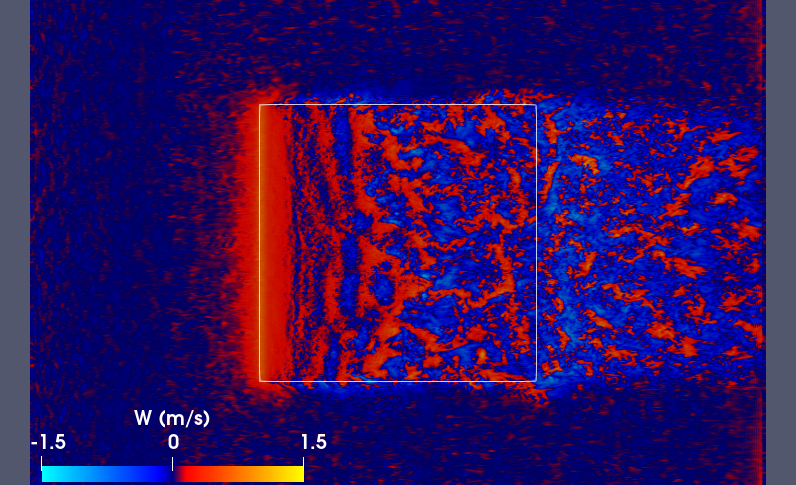}
        \caption{Snapshots of instantaneous w-velocity above the canopy showing areas of sweeps and ejections in the canopy for the Linn (top) and Smagorinsky (bottom) models at 1500 seconds. The white outline corresponds to the canopy and x and y are in m, w in $\frac{m}{s}$. }
        \label{paraviewW}
    \end{figure}

Figure \ref{paraviewU} shows the instantaneous streamwise velocity through the center of the canopy in the y-direction for both sets of simulations at 2,000 seconds after the simulations reached a quasi steady-state. Upwind of the canopy, the Linn model has developed significantly more turbulence compared to the Smagorinsky model. This was evident in the precursor runs as well, highlighting the difference between a full turbulent transport model (Linn) and an algebraic constitutive model (Smagorinsky). With both models, there are characteristic qualities of flow over a canopy, especially at the leading edge. When the flow runs into the leading edge, a positive vertical velocity distorts the flow and carries air upward towards the free-stream flow above the canopy. This is induced by the partial blockage of the flow provided by the thick vegetation and the associated positive pressure gradient (in streamwise direction) that forms due to the canopy drag (Pimont et al, 2022). Canopy drag and positive pressure gradient in the canopy drive a decrease in the streamwise velocity within the canopy, but also a slow down in the flow as it comes close to the leading forest edge and the flow accelerates above the canopy (Figure \ref{paraviewU}) (Pimont et al., 2022; Dupont and Brunet 2008, 2009; Pimont et al. 2009, Dupont et al., 2011). The vertical wind shear $\frac{\partial u}{\partial z}$ approaches zero and thus decreases the turbulence levels within the canopy and is nearly identical for both models because of the higher bulk density in the canopy. The mean wind profile through the canopy in both simulations are in agreement, both showing the effect of the canopy drag on the magnitude of the wind with a higher bulk density (Figure \ref{canopydrag}). The tree bulk density used in these simulations is representative of a typical bulk density for deciduous tree stands, such as Oak and Maple, with bulk density of $0.0974 \frac{kg}{m^3}$. 
  
A common phenomenon often discussed in fire-atmosphere interactions involving a canopy are sweeps and ejections. Vertical shear at the top of the canopy induces the development of eddies, which produce strong downdrafts (sweeps) into the canopy and weak updrafts (ejections) vertically out of the canopy, controlling most of the momentum and scalar transfer between the vegetation and atmosphere (Bebieva et al. 2021; Gao et al. 1989; Lu and Fitzjarrald, 1994). A comparison between the two-dimensional slices above the canopy of instantaneous vertical velocities, $w$, is seen in Figure \ref{paraviewW}. While the Linn model displays vertical movement of air all around the canopy compared to the Smagorinsky model, the vertical velocity above the canopy show areas of ejections and sweeps in both models. Through the canopy, the vertical movement of air in the Smagorinsky model occurs with an apparent pattern, consistent with the development of coherent structures. There is a characteristic positive vertical velocity adjacent to and at the leading edge, followed by the areas of updrafts and downdrafts downwind from the leading edge. Figure \ref{paraviewW} confirms this vertical velocity feature at the leading edge of the canopy, where positive vertical velocity accumulates up- and down-stream of the leading edge above the canopy.

\begin{figure}[h!]
        \centering
        \includegraphics[width=14cm]{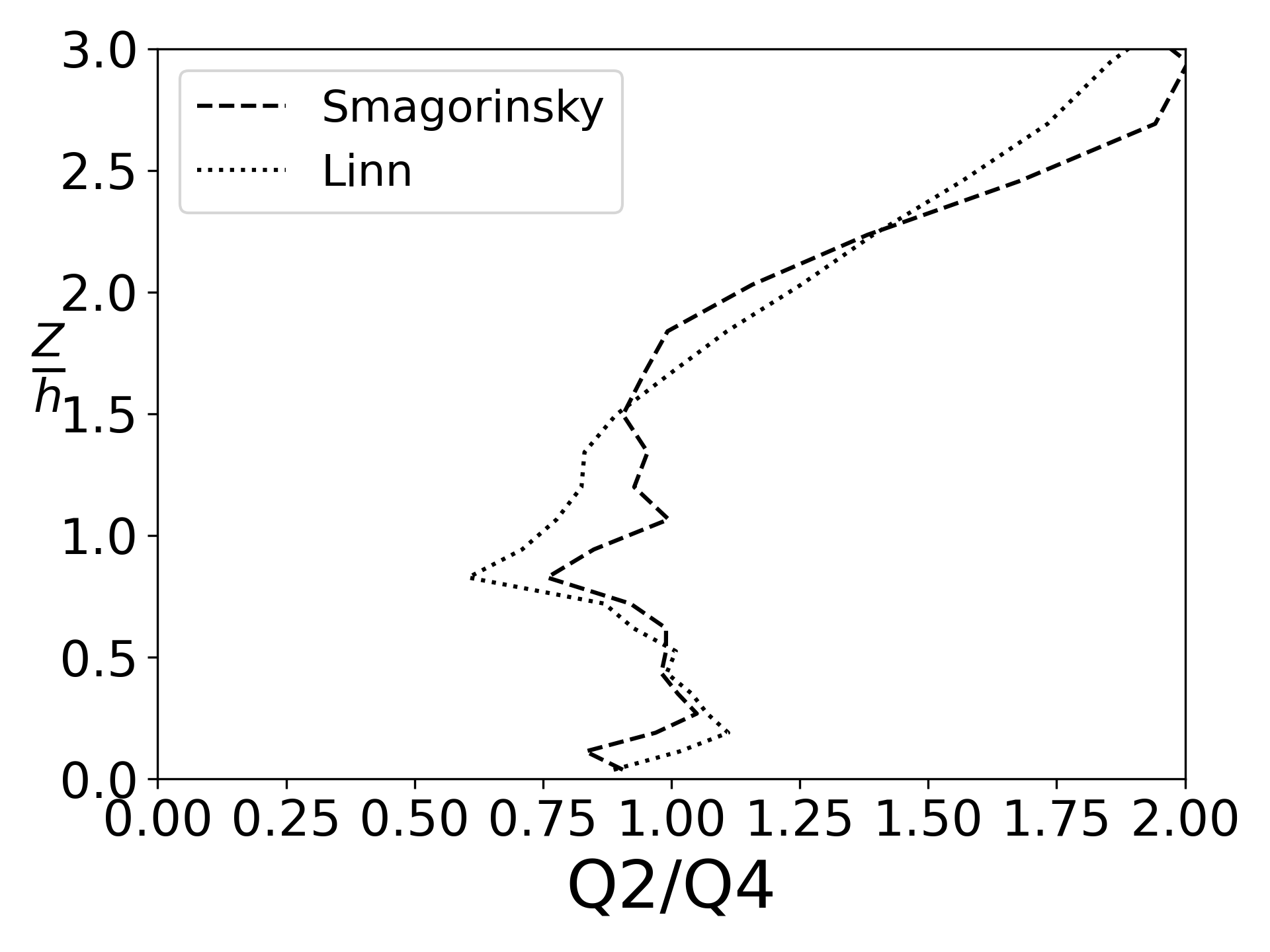}
        \caption{Time-averaged vertical profiles of the Q2/Q4 ratio of contributions to the Reynolds stress from the ejection quadrant (Q2) and the sweep quadrant (Q4) for both models.}
        \label{Q2Q4}
\end{figure}

The ratio between the Reynolds stresses, $\overline{u'w'}$, associated with ejections and sweeps allows the quantification of the overall relative contributions of these two eddy motions to the mean momentum flux between the slow moving winds within the canopy and those faster moving winds above the canopy (Dupont and Brunet, 2009). Velocity fluctuations can be decomposed into four quadrants. Quadrant 2 (Q2) ($u'<0,w'>0$) and quadrant 4 (Q4) ($u'>0,w'<0$) represent the areas of ejections and sweeps, respectively (Finnigan 2000). Studies have consistently found that the major contributor to momentum transfer within and just above the canopy is the ejections quadrant (Q2), representing updrafts of slow moving air out of the canopy, and the second most important is the sweep quadrant (Q4), representing downward gusts of high momentum air (Finnigan 2000). Time- and volume-averaged vertical profiles of the ejection/sweep Q2/Q4 ratio over the canopy from both simulations were computed and are presented in Figure \ref{Q2Q4}. As shown in previous LES data, in both models the sweep contribution is dominant within the canopy and just above the canopy, and the ejection contribution becomes more important farther above the canopy. This is consistent with what was observed in Katul et al. (2006) and Finnigan et al. (2009). A peak near the ground can be seen in both models where the ejections are dominating (Q2/Q4 $>$ 1). This is expected because these peaks occur at the top of the grass layer ($=$ 0.27m), where we would expect the ejection quadrant to start dominating.

\begin{figure}[h!]
        \centering
        \includegraphics[width=0.68\textwidth]{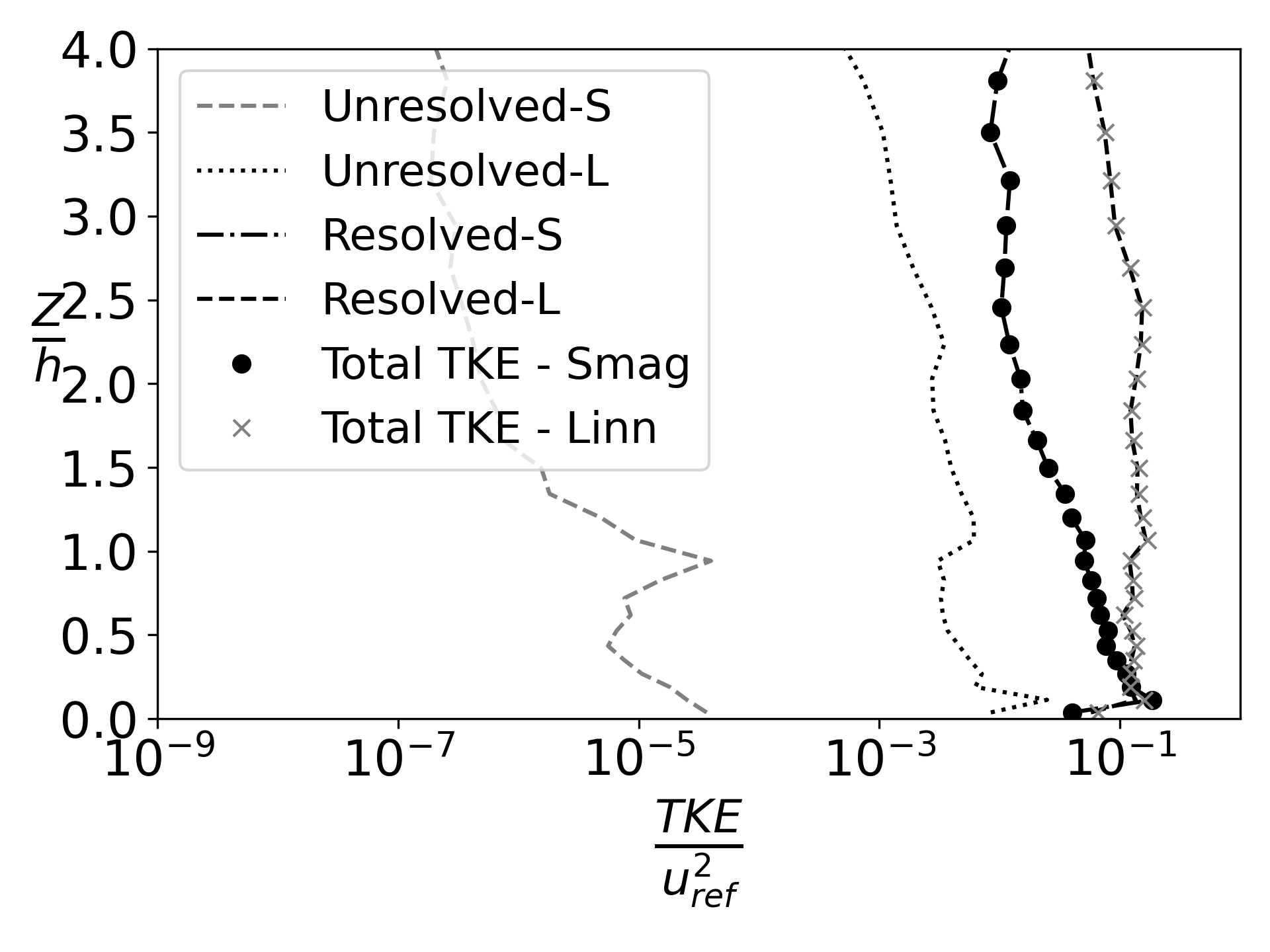} \\
        \includegraphics[width=0.68\textwidth]{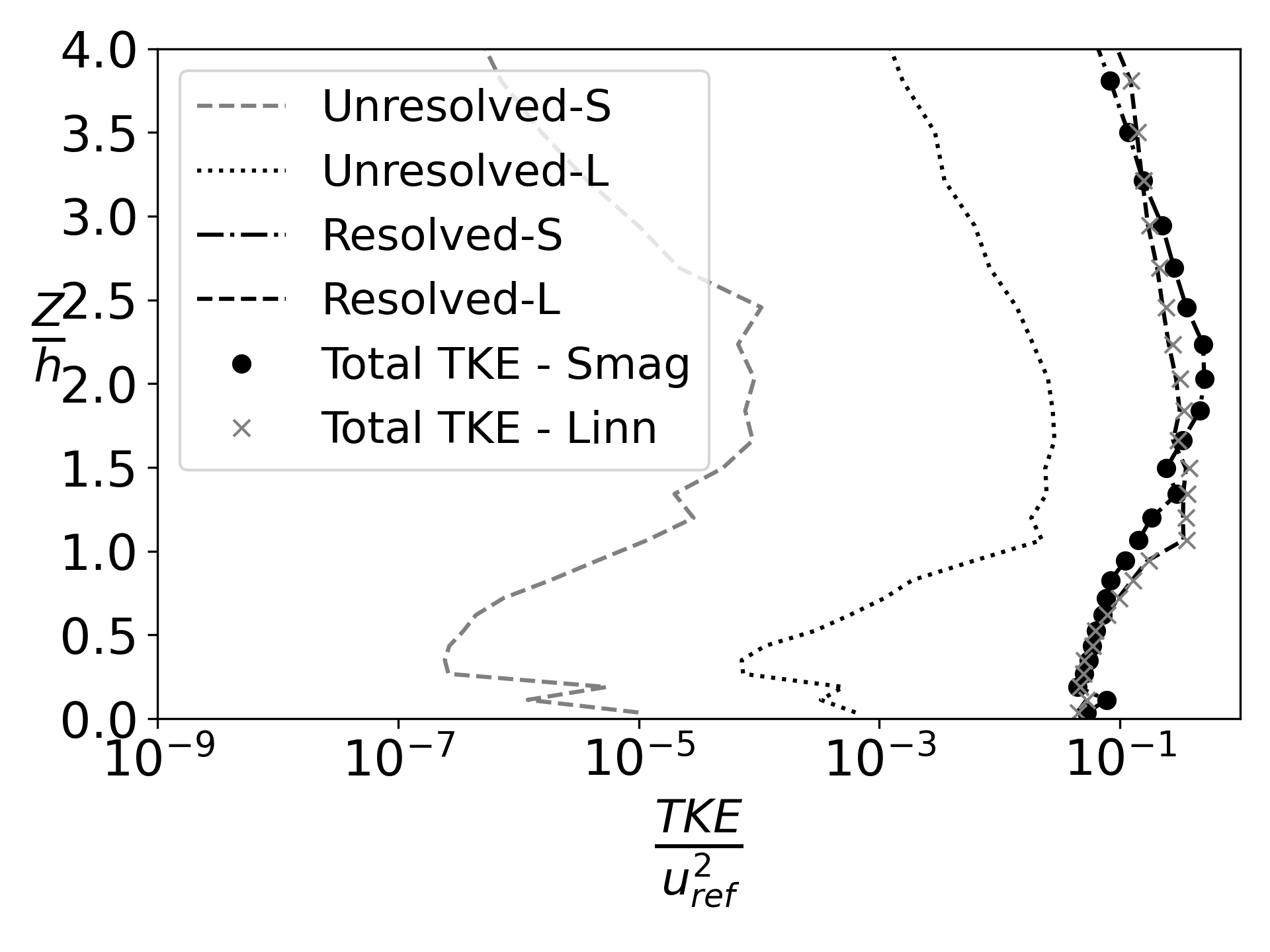}
        \caption{Time-averaged, normalized vertical profiles of TKE at the leading edge of the canopy (top) and through the center of the canopy (bottom) for both the Smagorinsky (-S) and Linn (-L) models showing contributions of energy from the unresolved and resolved scales.}
        \label{totalTKE}
\end{figure}

\begin{figure}[h!]
        \centering
        \includegraphics[width=7.5cm]{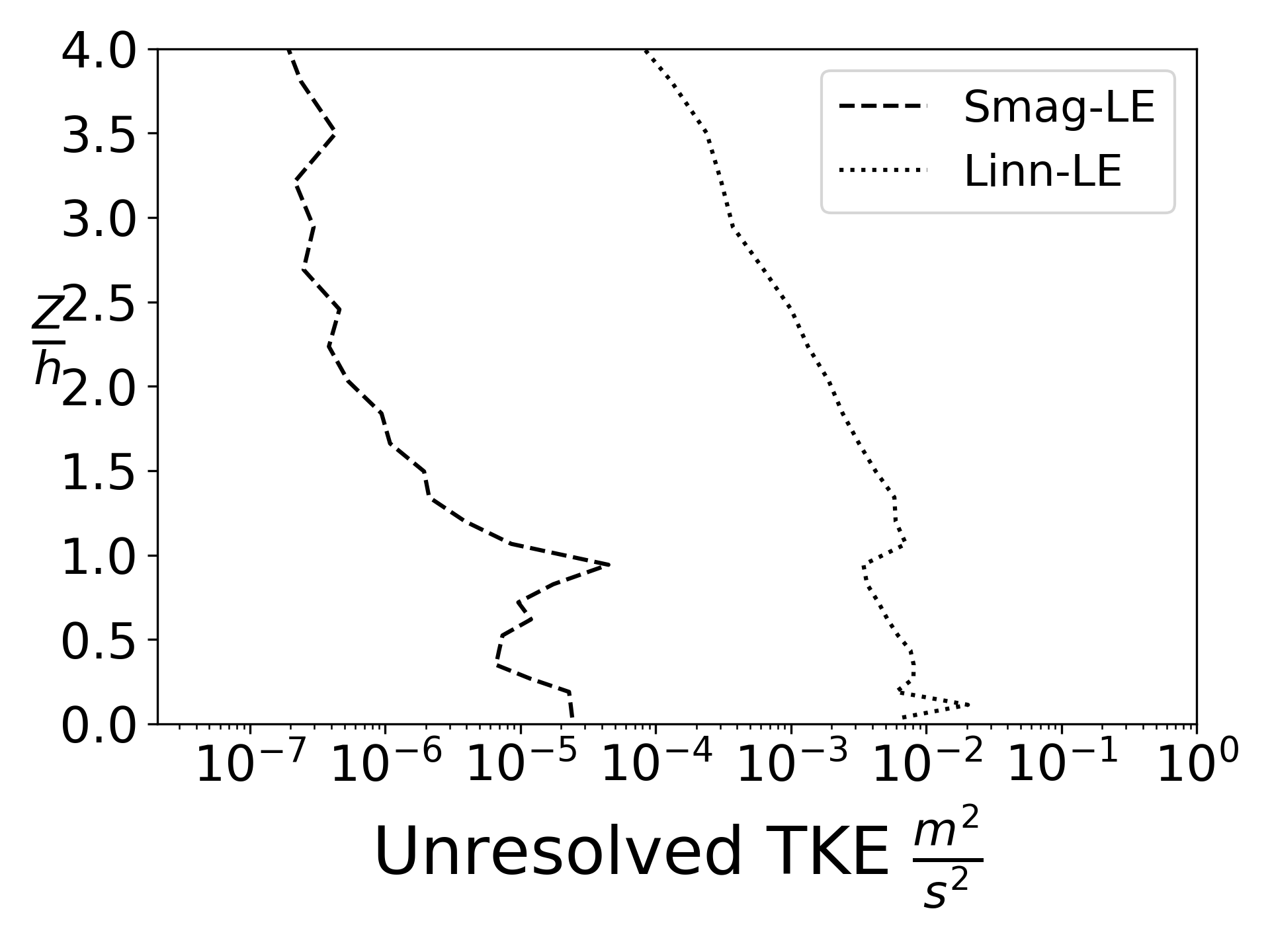}
        \includegraphics[width=7.5cm]{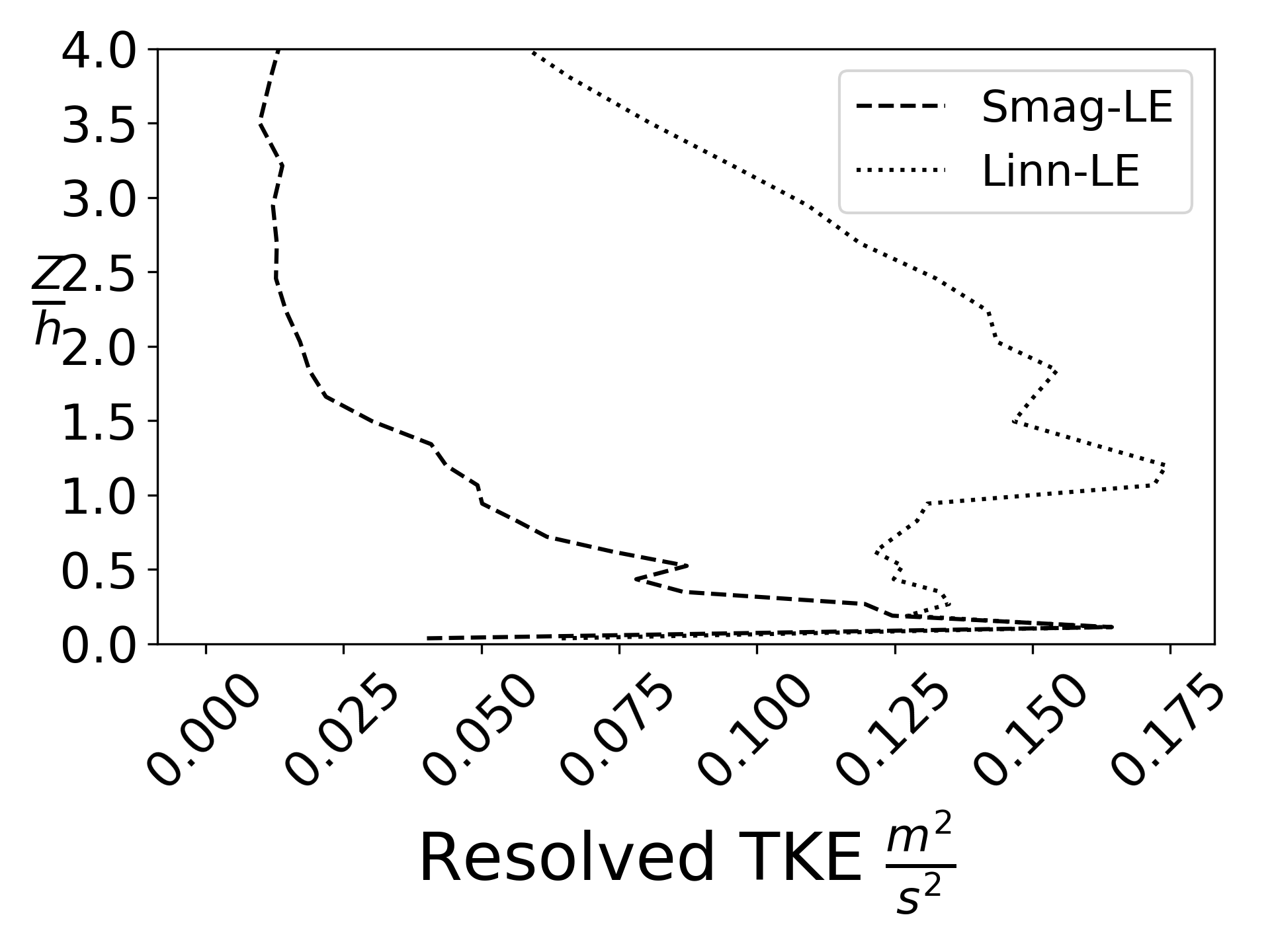}
        \includegraphics[width=7.5cm]{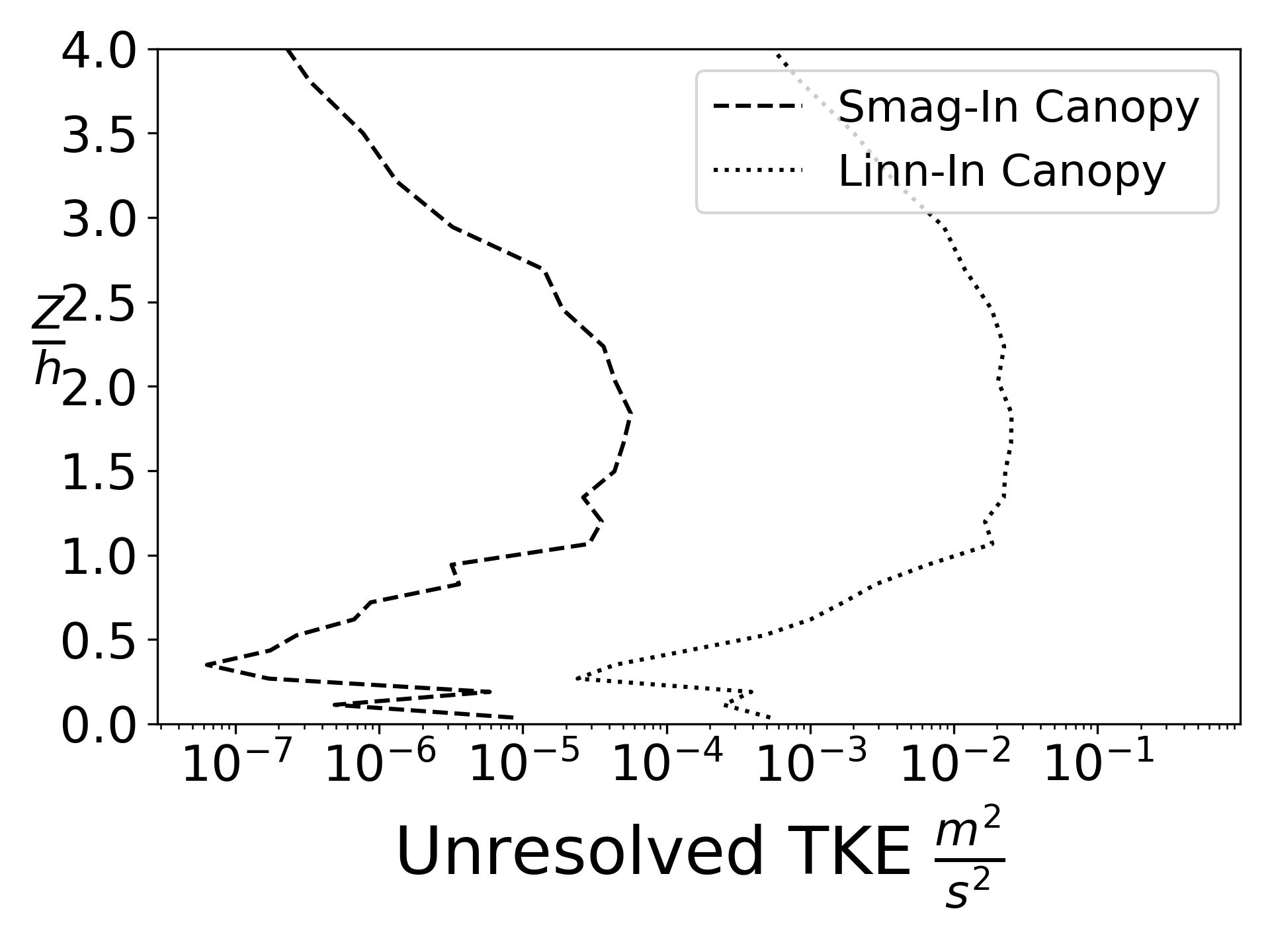}
        \includegraphics[width=7.5cm]{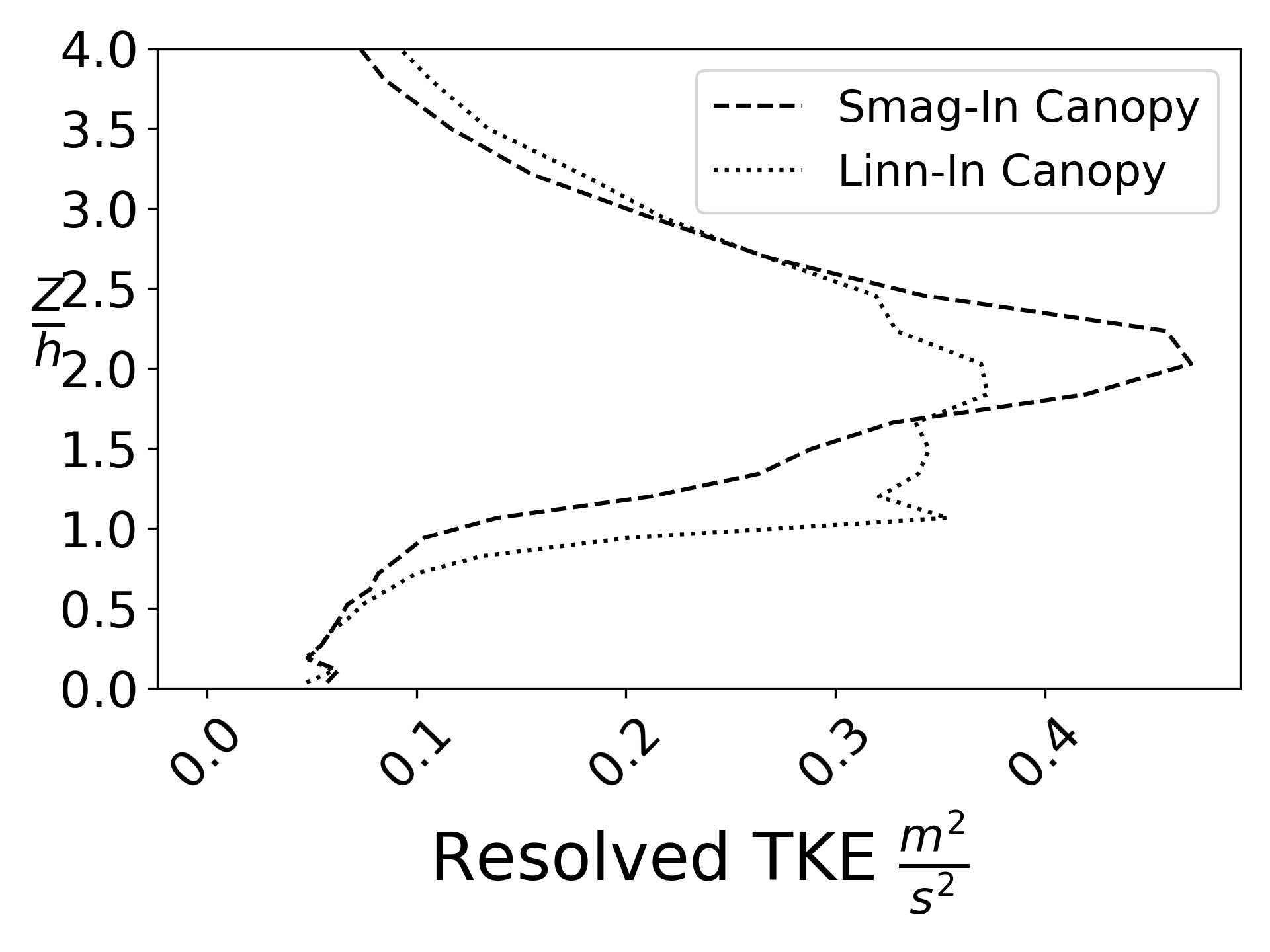}
        \includegraphics[width=7.5cm]{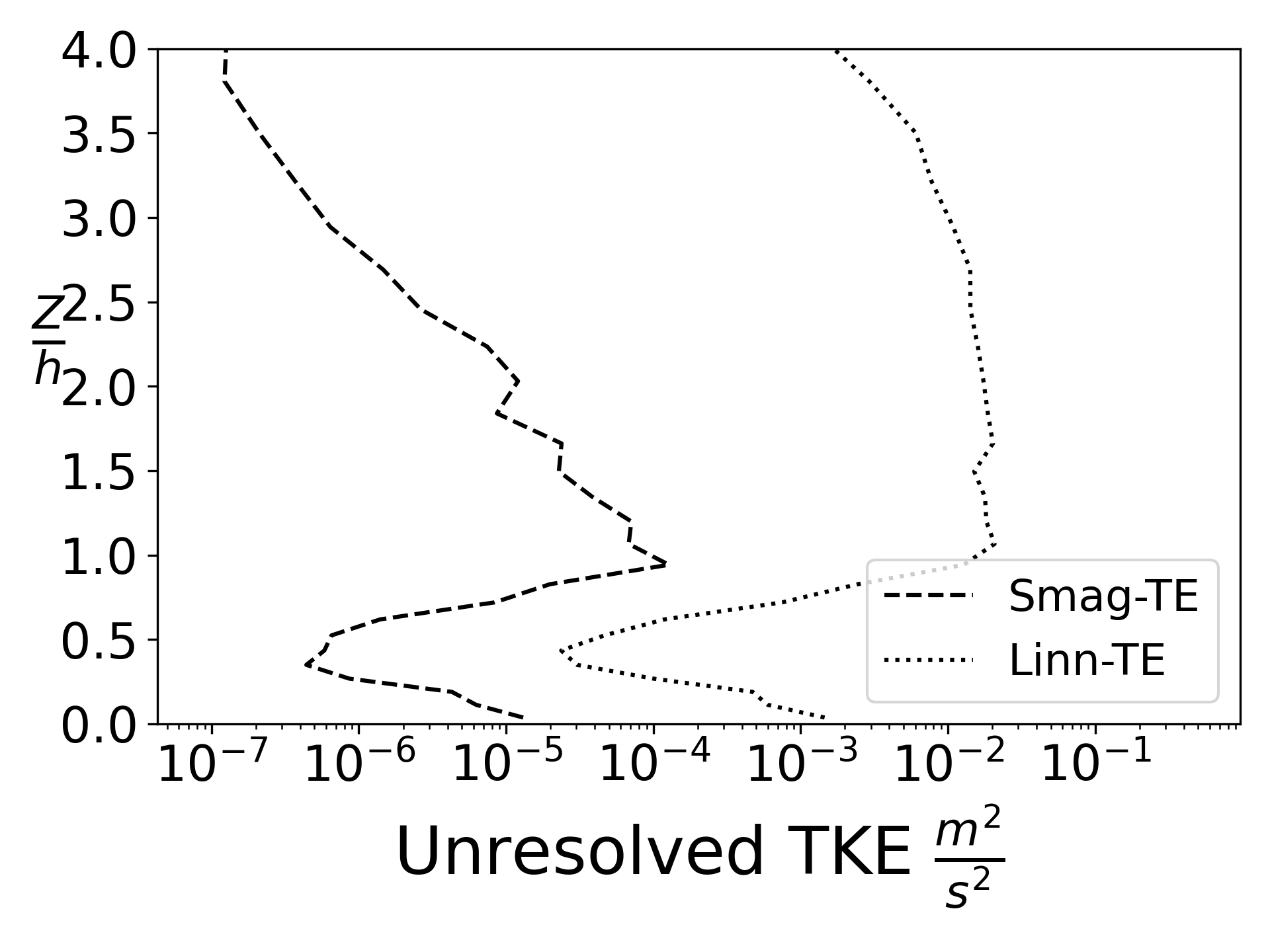}
        \includegraphics[width=7.5cm]{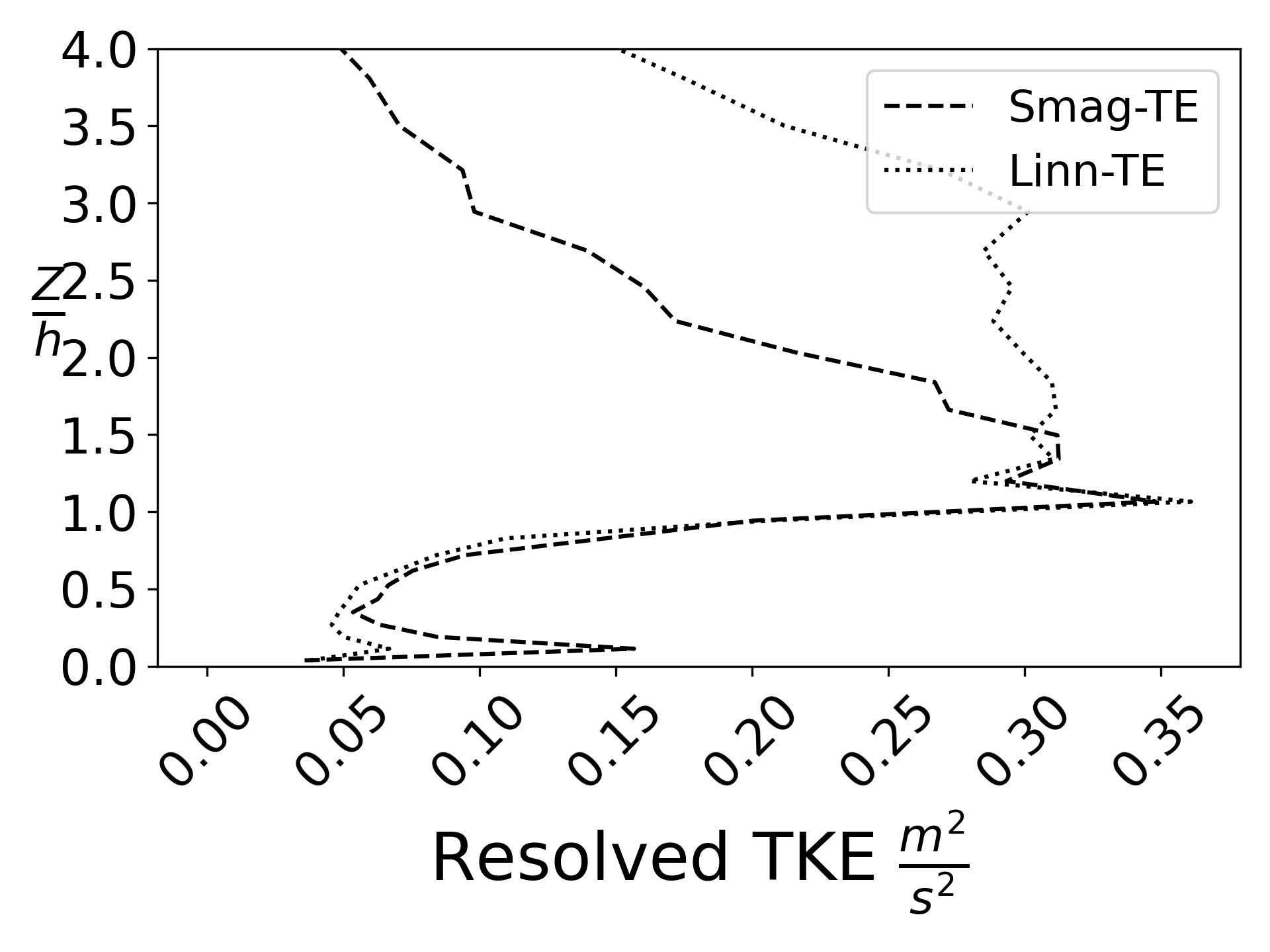}
        \caption{Unresolved ($\langle k_A \rangle_t$ and $\langle k_S \rangle_t$) and resolved TKE ($0.5 \langle \tilde{u}'_i \tilde{u}'_i \rangle_t$) comparison of the Linn and Smagorinsky models at the leading edge of the canopy (-LE), in the canopy (-In Canopy), and the trailing edge of the canopy (-TE).}
        \label{unresolved}
\end{figure}

\newpage
The total TKE is composed of one resolved and three modeled parts ($K_A$, $K_B$, $K_C$) in the Linn turbulence model. $K_A$ represents the largest unresolved scale in FIRETEC, and the Smagorinsky model contains one unresolved scale, which was chosen to be the same as the $K_A$ scale. Thus, the comparisons of the unresolved TKE are done with the unresolved TKE for the Smagorinsky model and $K_A$. The adequacy of a mesh for LES can be inferred from the examination of the resolved and unresolved stress components (Mirocha et al. 2010). Figure \ref{totalTKE} shows vertical profiles of turbulent kinetic energy for both models at both the leading edge of the canopy and in the center of the canopy. It is clear that for this study, the contributions of the modeled parts (unresolved) are significantly smaller than the resolved TKE, representing less than 5\% of the total TKE, as has been seen in previous studies (Pimont et al. 2009). Pimont et al. (2009) showed that for simulations without fire and under neutral conditions, the decomposition of the SGS TKE into three frequency bands is not necessary. These results are consistent with those of Shaw and Patton (2003), who showed with LES that smaller scales enhance dissipation of subgrid-scale energy and can simply be represented through an increase of the cascade term in an SGS TKE conservation equations (Pimont et al. 2009). Thus, the formulation of the Smagorinsky eddy viscosity model in this application suffices.

Figure \ref{unresolved} shows direct comparisons of the time averaged vertical profiles of unresolved TKE for both models and the resolved TKE for both models at different locations in the domain: the leading edge (LE), through the center of canopy (In Canopy), and at the trailing edge (TE). The unresolved TKE for the Linn model, at every location in the domain, is larger than that of the Smagorinsky model by orders of magnitude. However, they both share the same qualitative profile shapes. As described above, the amount of turbulence is a feature of the Linn model which carries persistent dynamic variables and permits backscatter, whereas the Smagorinsky model has neither. It comes as no surprise that the Smagorinsky model produces less modeled TKE.

The progression from the leading to the trailing edge in the resolved TKE plots show that the canopy has a dramatic effect on the Smagorinsky TKE profiles. At the leading edge, the Smagorinsky resolved turbulence was largely developed over grass in the precursor simulations and this is shown by the rapid decay of the profile to background with height from the surface. The resolved turbulent layer is deeper in the Linn model. At the middle of the canopy, the resolved turbulence has developed significantly with a peak at $z/h \approx 2$ for the Smagorinsky model. This profile shows that a turbulent boundary layer has developed over the canopy and, qualitatively, it compares well with the Linn profile. In particular, the portions of the profile near the ground and at $z/h > 2.5$ have very good qualitative agreement. At the trailing edge, the profile within the canopy is nearly identical for both models with a higher peak for the Smagorinsky model at the top of the grass layer. Above the canopy at the trailing edge, the Smagorinsky TKE decays to background with height much lower than the Linn model, which is still developing turbulence due to the additional aforementioned mechanisms of that model.


\begin{figure}[h!]
    \begin{minipage}[t]{.65\linewidth}
        \centering
        \includegraphics[width=0.9\textwidth]{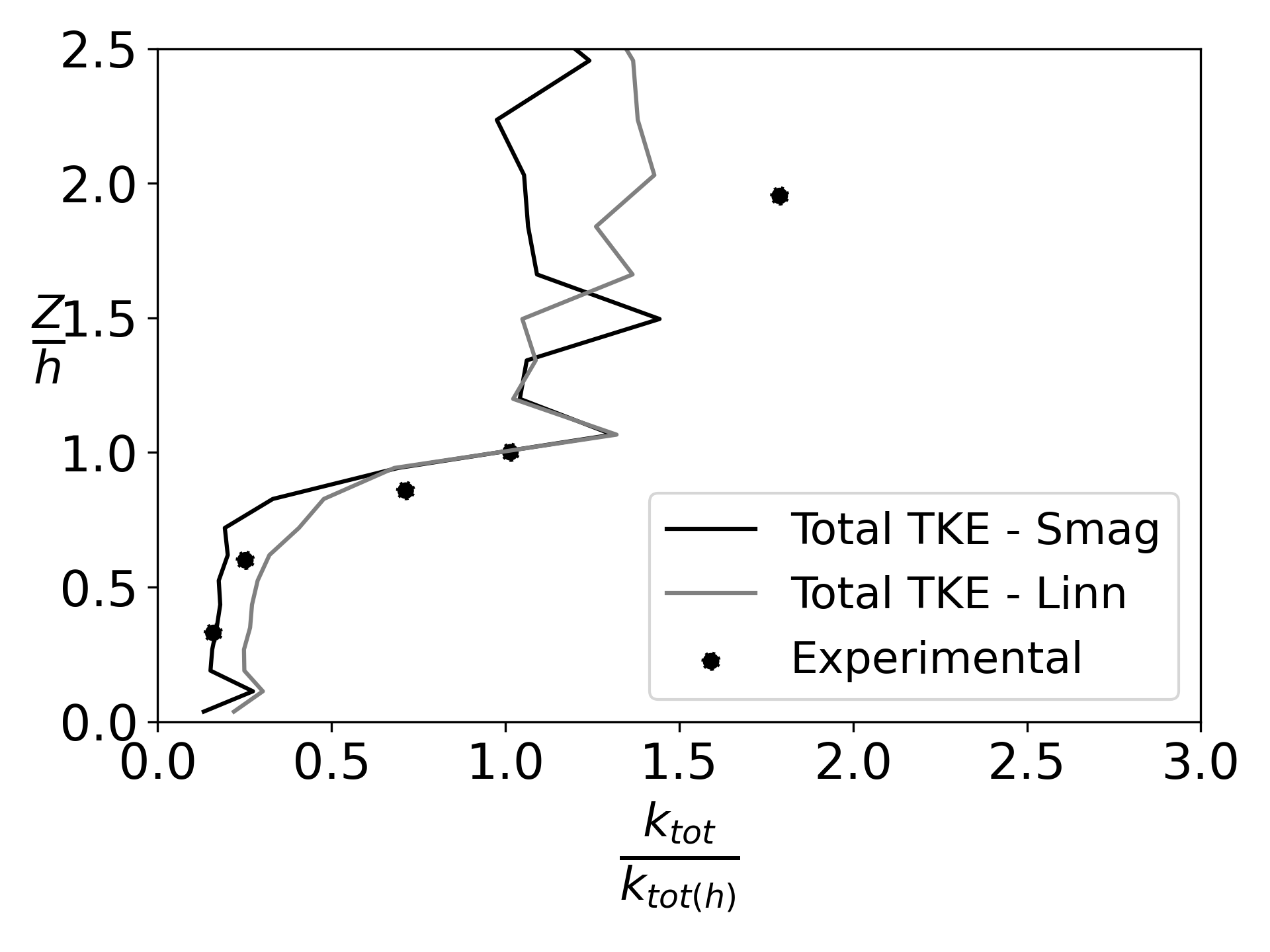}
    \end{minipage} \hfill
    \begin{minipage}[t]{0.45\linewidth}
    \centering
    \vspace{-2in}
    \hspace{-0.7in}
    \begin{tabular}{c c}
    \hline
       $RMSE$ for Smagorinsky model & 0.57\\
       $RMSE$ for Linn model  & 0.35 \\
    \hline
    \end{tabular}
    \label{experimental}
    \end{minipage}
    \caption{Comparison of the Linn model, Smagorinsky model, and field observations from Shaw et al. (1988) in a homogeneous forest canopy of normalized time-averaged vertical profiles of total turbulent kinetic energy $k_{tot}$. The simulated (lines) profiles were taken within the canopy at the $x,y$ point $(400,400)$ to better compare to the continuous canopy conditions in Shaw et al. (1988). The dataset was taken from Pimont et al. (2009). Corresponding root mean square errors are tabulated with experimental data from Shaw et al. (1988) as the actual value and the simulated normalized TKE for both models as the predicted values.}
    \label{experimental}
\end{figure}

Figure \ref{experimental} compares the mean vertical profile of the simulated total TKE for the Smagorinsky model and the Linn model with observed data from Shaw et al. (1988). This comparison was done by digitizing the normalized data points from Pimont et al. (2009). Compared to observations, both models performed reasonably well. Within the canopy, the profile is characterized by a decay of TKE for both models. It is important to note here that the bulk density of the canopy induces strong canopy drag. At $z/h \approx 2$, the normalized TKE is underpredicted in both models. Differences are expected because the experimental data was in a deciduous, heterogeneous, continuous forest, whereas the simulations are done with a deciduous, homogeneous forest. Nonetheless, both models compare well with experiment inside the canopy.

    \begin{figure}[h!]
        \centering        
        \includegraphics[width=0.75\textwidth]{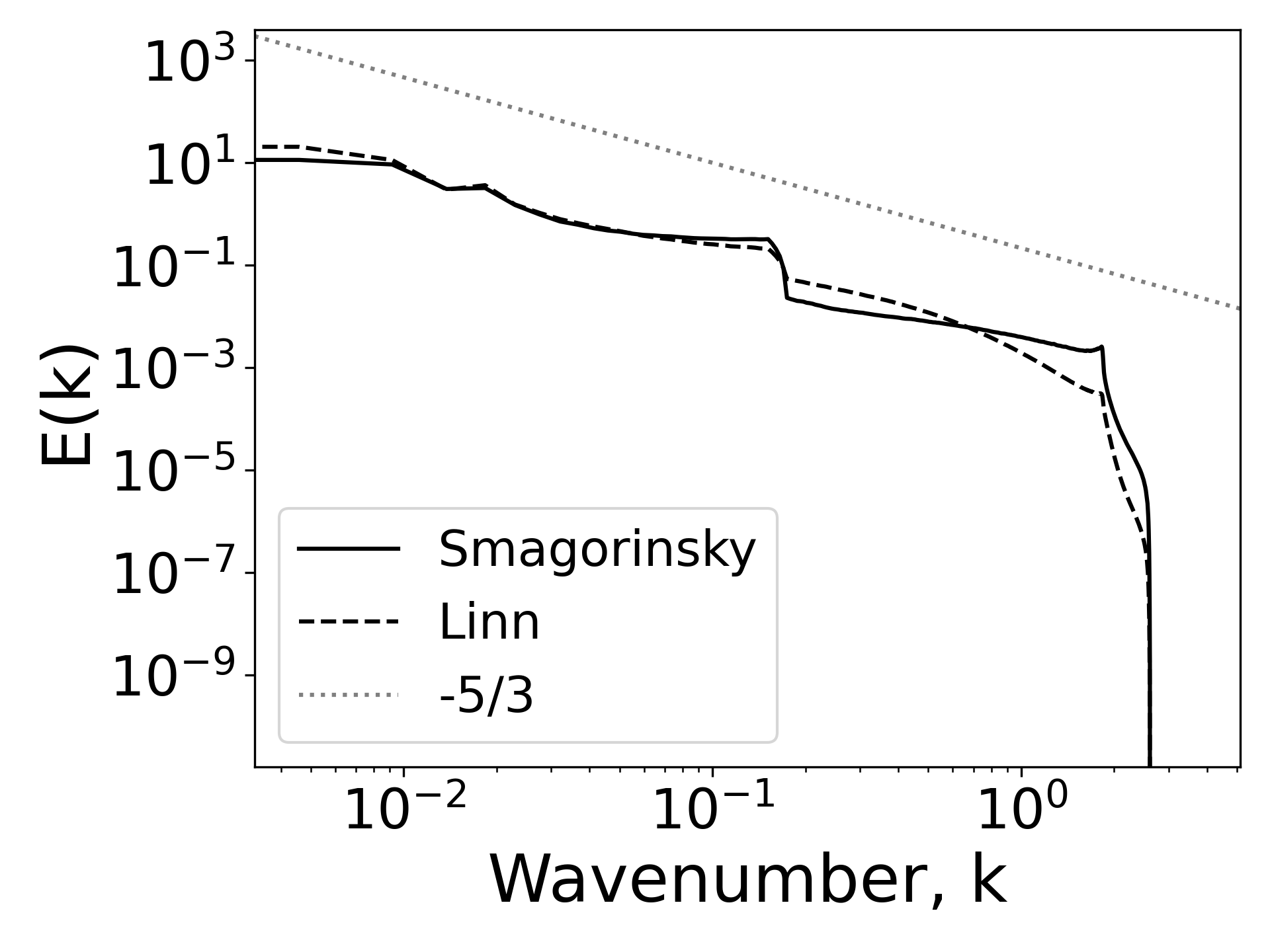}
        \vspace{-0.3in}
        \caption{Spectra of time- and volume-averaged kinetic energy of the velocity field above the canopy for the Smagorinsky model and the Linn turbulence model. The $-5/3$ slope from the Kolmogorov power law is plotted for reference in the inertial subrange.}
        \label{spectra}
    \end{figure}

For a successful LES result, we expect to resolve the largest, most energetic scales of turbulence in the flow, which follows when the filter width is well within the inertial subrange (Mirocha et al., 2010). Yet, the presence of a canopy significantly affects the turbulence spectrum at small scales (Pimont et al. 2022). Spatial spectra are often used to quantify the proportion of different length scales occurring in a turbulent flow (Mirocha et al. 2010). Figure \ref{spectra} shows the time- and volume-averaged spectra of kinetic energy of the velocity field for both models just above the canopy. The mesh size chosen for this study produced the  $-5/3$ slope of the inertial subrange scales for the Smagorinsky and Linn simulations as expected, according to the Kolmogorov power law. The smaller energy-producing turbulent structures are captured well. There is a feature for $0.1 < k < 0.2$ in the inertial subrange, where energy is injected from the turbulent boundary layer. This energy injection separates the large-scale motions, that follow the $-5/3$ power law at lower wave numbers, and the small, dissipation scales at higher wave numbers, known as the enstrophy subrange. The Linn model in this subrange follows the power law for the enstrophy inertial range with a $-3$ slope for $0.4 < k < 2.0$. This feature is due to  back scatter from high to low wave numbers. As stated above, the Smagorinsky model does not account for this mechanism, and thus does not have an enstrophy subrange between the $-5/3$ inertial subrange and the dissipation scales. Instead, the Smagorinsky model more-or-less maintains a $-5/3$ slope to the dissipation scales. In Figure \ref{spectra}, there is a sharp increase in energy near the grid scale where energy is not sufficiently dissipated. The grid scale peak is more pronounced in the Smagorinsky model. This pattern is consistent with the findings of Guimond et al. (2016) for HIGRAD, the atmospheric dynamics model that is coupled to FIRETEC. This peak was attributed to a build-up of small-scale energy and was shown to disappear after a hyper-viscosity was implemented. This hyper-viscosity was not utilized in this study.

\section*{Discussion}

Large coherent turbulent structures above the canopy develop in 3 stages: the primary Kelvin-Helmholtz instability during the shear, clumping into rollers connect by broad regions of highly strained fluid, followed by secondary instabilities in the rollers (Pimont et al. 2002; Finnigan 2000). The size of these coherent structures has been found to be on the order of 10's of meters (Pimont et al., 2022). These structures are explicitly resolved and account for the majority of the TKE, as can be seen in the results presented above. Comparing the resolved TKE in both model implementations became important to establish the reliability of the hydrodynamics in both models. We examined the most turbulence-producing locations in the domain, namely the edges of the canopy and above of the canopy midway across the stand, and found that the smaller, SGS turbulent structures account for less than 5\% of the total TKE. These smaller structures had relatively larger TKE values at the canopy height and near the ground for both the Linn and Smagorinsky models. In the Linn model, the increase in energy of eddies and the turbulent scale peak comes from the production of turbulence associated with shear, while the viscous dissipation induces a decrease in energy of eddies with smaller size (Pimont et al. 2022).

Some differences captured in this study highlight that the Linn model produces more turbulence, specifically unresolved TKE, with a difference of two orders of magnitude in the unresolved TKE between the models. The Linn model has two evolution equations for the TKE at two length scales and these evolve the sources and sinks. In other words, the Linn unresolved TKE is dynamic and persists in time, whereas the Smagorinsky TKE does not. The higher levels of TKE in the Linn model are more representative of canopy flow than the Smagorinsky model because it models physically relevant processes in the canopy, such as drag, at various scales. The Smagorinsky model relies on resolved shear to dissipate energy through a viscous process at a single length scale, $\Delta$.

The Linn model carries three frequency bands of turbulent energy whose models are tailored to flow through vegetation. The turbulent kinetic energy at the largest unresolved scale, $k_A$, is evolved with a conservation equation with sources and sinks that cascade to smaller scales. The time dependent evolution and spatial transport of $k_A$ is consistent with the evolution of the governing mass, momentum and energy equations. This additional fidelity in the Linn model permits a time history and non local influences of turbulence that is regularized by advection, scale dependent shear production; destruction driven by drag and dissipation; and viscous redistribution. The Linn model also has the ability to backscatter unresolved energy to the resolved scales as shown in Figure \ref{spectra}. Thus, it is expected that there is more turbulence in the Linn model than in the Smagorinsky model. These differences highlight the fact that the Smagorinsky model computes a turbulent viscosity that reflects shear at that moment and location with no dependence on local history or upstream flow characteristics. The Smagorinsky turbulent viscosity is proportional to the local velocity gradients and it directly diffuses the velocity gradient, reducing the shear. This action is repeated every time cycle with a new snapshot of viscosity, determined by the instantaneous velocity gradients. Furthermore, the Smagorinsky model does not have a mechanism to dissipate energy due to drag on the vegetation. Instead, the vegetation drag from the momentum equation leads to resolved shear that manifests as turbulent viscosity.

In the turbulence spectra, as the length scales approach that of the mesh size, it becomes clear that neither model dissipates enough energy to smaller scales. The Smagorinsky model does not carry a dissipation term to serve as an energy sink. The Linn model considers a dissipation term in both transport equations for $k_A$ and $k_B$ which transfer energy to smaller scales and it does a better job with dissipation. The difference in the dissipation to small scales can be seen in Figure \ref{unresolved}, where the unresolved TKE in the Linn model is much larger.

These results show that the presence of a canopy with exposed edges (surrounded by fuel breaks) significantly impacts the wind field and modifies the dynamics observed. This is expected given the strong influence of the canopy drag, which dampens the momentum as the wind moves through the canopy. Although the unresolved TKE values from these two modeling approaches are substantially different, the similarities of the mean flow patterns and profiles suggest that the simpler Smagorinsky model can be substituted into the FIRETEC model for dense and relatively homogeneous forest canopy scenarios without fire for further analysis of the mean flow. Such further analysis might be for example, a study of the vorticity budget in these scenarios.



\section*{Conclusions}
Understanding the role of wind dynamics around canopies is an important part of comprehensive numerical modeling fire-atmosphere interactions. In order to enable deeper analysis into the phenomena associated with wind interaction with canopies, it is advantageous to have a simpler closure scheme for unresolved turbulence than what is currently implemented in HIGRAD/FIRETEC. However, the criteria for substituting a simpler closure scheme is that the characteristics of the flow field are not substantially different from those produced by the previously validated FIRETEC model, which uses a three-scale TKE model with multiple transport equations for turbulence.  

In this study, the Smagorinsky eddy viscosity model was compared to the Linn turbulence model currently implemented HIGRAD/FIRETEC. The results of this investigation serve to show that HIGRAD/FIRETEC can produce comparable resolved flow fields using Smagorinsky eddy viscosity model. While the HIGRAD/FIRETEC model produced more turbulence (and TKE) overall, the resulting resolved flow characteristics are fairly similar in comparing profiles in and around the canopy. Results are also comparable with those from experimental data of turbulent statistics through a canopy. The contours of sweeps and ejections also reveal that these two models produced similar results in the way momentum is transferred from above into the canopy. Unresolved TKE is higher in the Linn turbulence model. This is due to the fact that TKE is transported through time and space in the Linn model and upstream production of turbulence can affect downstream locations. The mesh size used in the Smagorinsky and Linn simulation were in agreement with the $-5/3$ slope of the inertial subrange scales that are expected from the Kolmogorov power law and the shape of the spectrum was consistent with previous studies of the atmospheric hydrodynamics model coupled to FIRETEC.

Results of this study show that the Smagorinsky model produces similar representation of the resolved flow field around a canopy. This implementation will be used in Part 2 of this study, which involves deriving a vorticity budget equation from the momentum equation in HIGRAD/FIRETEC. Effectively, the turbulent residual stress tensor, represented as the Reynolds stress, is replaced by the residual stress tensor for the Smagorinsky eddy viscosity model. This facilitates the derivation and overall analysis of the vorticity budget, since the Linn turbulence model carries conservation equations for three turbulent frequencies. Findings from the vorticity budget analysis will follow in Part 2 of this investigation.

HIGRAD/FIRETEC's ability to simulate wind-flow dynamics in neutral conditions was validated by Pimont et al. (2009). They showed that ``modifications of the wind-flow velocity and turbulence induced by the presence of a break can be very well described by the physically-based model FIRETEC" (Pimont et al. 2009). Well-studied and common concepts of wind flow in and around a canopy were observed in our study. These included the influence of canopy drag on the mean flow, the momentum transfer by sweeps and ejections just above the canopy, and the effects of wind shear near the canopy. Understanding the way turbulent flow interacts with a canopy, and vice versa, in these wind-only simulations is the first step in investigating the full scope of fire-atmosphere interactions.

In order to investigate additional aspects of the wind-canopy interaction, namely vorticity, additional analysis using tools such as deriving a vorticity budget equation is valuable.  However, the Linn formulation with additional transport equations for multiple portions of the TKE spectrum makes this additional analysis difficult.  Thus, it is desirable to substitute a simpler closure scheme into the model if the resultant flow patterns are similar. This is the motivation for the present study investigating whether the flow fields are sufficiently similar between implementations of the local Smagorinsky turbulence scheme and the Linn turbulence scheme. Introducing a heat source or ignition would add to the complexity of these simulations due to the introduction of plume dynamics. The influence of fire requires further investigation and is beyond the scope of this study. Although HIGRAD/FIRETEC has been validated against experimental data, understanding the way turbulence is handled by direct comparisons with an implementation of a well-studied turbulence model is extremely useful. Wind flow, the presence of a canopy, and turbulence dominate fire behavior, thus further exploring these dynamics helps provide a better understanding of the way they interact with a fire.
\\

\noindent \textbf{Dorianis M. Perez:} Conceptualization, Methodology, Software, Formal Analysis, Writing - Original Draft, Visualization, Writing - Review \& Editing; \textbf{Jesse Canfield:} Conceptualization, Writing - Review \& Editing, Supervision; \textbf{Rodman Linn:} Writing - Review \& Editing, Funding Acquisition; \textbf{Kevin Speer:} Conceptualization, Writing - Review \& Editing 

\section*{Acknowledgements}
This work was supported by the U.S. Department of Energy through the Los Alamos National Laboratory. Los Alamos National Laboratory is operated by Triad National Security, LLC, for the National Nuclear Security Administration of U.S. Department of Energy (Contract No. 89233218CNA000001). We express our sincere gratitude to the U.S. Department of Defense, Strategic Environmental Research and Development Program (SERDP) for partially funding this study under grant RC20-C3-1298. We additionally thank Los Alamos National Laboratories Center for Space and Earth Science (CSES) for partially funding this work under LDRD project 20240477CR-SES. All computations for this work were performed using Los Alamos National Laboratory's Institutional Computing Resources. 

\section*{References}

\hspace{0.2in} 
Bebieva, Y., Speer, K., White, L., Smith, R., Mayans, G., Quaife, B. Wind in a natural and artificial wildland fire fuel bed. Fire, 4, 30, 2021. \\

Banerjee, T., Heilman, W., Goodrick, S., Hiers, J. K., Linn, R. Effects of canopy midstory management and fuel moisture on wildfire behavior. Scientific reports, 10(1), 17312. 2020. \\

Canfield, J.M., Linn, R.R., Cunningham, P., Goodrick, S.L. Modelling effects of atmospheric stability on wildfire behaviour. In ‘Proceedings 6th Fire and Forest Meteorology Symposium and 19th Interior West Fire Council Meeting’, 25–27 October 2005, Canmore, AB, Canada. \\

Cheney, N.P., Gould, J.S., Catchpole, W.R. Prediction of fire spread in grasslands. International Journal of Wildland Fire 8(1), 1–13, 1998. \\

Colman, J. J., Linn, R. R. Separating combustion from pyrolysis in HIGRAD/FIRETEC. International Journal of Wildland Fire, 16 (4), 493–502, 2007. \\

Dupont, S, Bonnefond, J-M, Irvine, MR, Lamaud, E, Brunet, Y. Long-distance edge effects in a pine forest with a deep and sparse trunk space: In situ and numerical experiments. Agricultural and Forest Meteorology 151(3), 328–344, 2011. \\

Dupont, S., Brunet, Y. Edge flow and canopy structure: a large-eddy simulation study. Boundary-Layer Meteorology 126, 51–71, 2007. \\

Dupont, S., Brunet, Y. Influence of foliar density profile on canopy flow: a large-eddy simulation study. Agricultural and Forest Meteorology 148, 976–990, 2008a. \\

Dupont, S, Brunet, Y. Coherent structures in canopy edge flow: A large-eddy
simulation study. Journal of Fluid Mechanics 630, 93–128, 2009. \\

Finnigan, J. Turbulence in plant canopies. Annual Review of Fluid Mechanics 32, 519–571, 2000. \\

Finnigan, J.J., Shaw, R.H., Patton, E.G. Turbulence structure above a vegetation canopy. Journal of Fluid Mechanics 637, 387-424, 2009. \\

Foudhil, H., Brunet, Y., Caltagirone, J.P. A fine-scale $\kappa–\epsilon$; model for atmospheric flow over heterogeneous landscapes. Environmental Fluid Mechanics 5(3), 247–265, 2005. \\

Gao, W., Shaw, R.H., Paw, U.K.T. Observation of organised structures in turbulent flow within and above a forest canopy. Boundary-Layer Meteorology, 47(1), 349–377, 1989. \\

Green, S.R. Modelling turbulence air flow in a stand of widely spaced trees, PHOENICS. Journal of Computational Fluid Dynamics 5, 294–312, 1992. \\

Guimond, S.R., Reisner, J.M., Marras, S., Giraldo, F.X. The impacts of dry dynamic cores on asymmetric hurricane intensification. Journal of the Atmospheric Sciences, 73(12), 4661-4684, 2016. \\

Hernandez, C., Keribin, C., Drobinski, P., Turquety, S. Statistical modelling of wildfire size and intensity: A step toward meteorological forecasting of summer extreme fire risk. Annales Geophysicae 33(12), 1495–1506, 2015. \\

Horiuti, K., Tamaki, T. Nonequilibrium energy spectrum in the subgrid-scale one-equation model in large-eddy simulation. Physics of Fluids 25(12): 125104, 2013. \\

Josephson, A. J., Casta$\tilde{n}$o, D., Koo, E., Linn, R. R. Zonal-Based Emission Source Term Model for Predicting Particulate Emission Factors in Wildfire Simulations. Fire Technology, 2020. \\

Kanda, M., Hino, M. Organized structures in developing turbulent- flow within and above a plant canopy, using a LES. Boundary-Layer Meteorology 68(3), 237–257, 1994. \\

Katul, G. G., Poggi, D., Cava, D., Finnigan, J. J.  The relative importance of ejections and sweeps to momentum transfer in the atmospheric boundary layer. Boundary-Layer Meteorology 120, 367–375, 2006. \\

Koo, E., Pagni, P., Weiss, D., Woycheese, J. Firebrands and spotting ignition in large-scale fires. International Journal of Wildland Fire, 19(7), 818-843, 2010. \\

Koo, E., Linn, R.R., Pagni, P.J., Edminster, C.B. Modelling firebrand transport in wildfires using HIGRAD/FIRETEC. International Journal of Wildland Fire, 21(4), 396-417, 2012. \\

Li, Z., Lin, J.D., Miller, D.R. Air flow over and through a forest edge: a steady-state numerical simulation. Boundary-Layer Meteorology 51, 179–197, 1990. \\

Linn, R.R. A transport model for prediction of wildfire behavior. Los Alamos National Laboratory, Science Report LA-13334-T 1997. (Los Alamos, NM) \\

Linn, R. R., Reisner, J. M., Colman J. J., Winterkamp J. Studying wildfire behavior using FIRETEC. International Journal of Wildland Fire, 11, 233–246, 2002. \\

Linn, R.R., Cunningham P. Numerical simulations of grass fires using a coupled atmosphere–fire model: basic fire behavior and dependence on wind speed. Journal of Geophysical Research 110, D13107, 2005. 

Linn, R. R., Winterkamp, J. L., Colman, J. J., Edminster, C., Bailey, J. D. Modeling interactions between fire and atmosphere in discrete element fuel beds. International Journal of Wildland Fire, 14(1), 37–48, 2005. \\

Linn, R. R., Anderson, K., Winterkamp, J. L., Brooks, A., Wotton, M., Dupuy, J.-L., . . . Edminster, C. Incorporating field wind data into FIRETEC simulations of the International Crown Fire Modeling Experiment (ICFME): preliminary lessons learned. Canadian Journal of Forest Research, 42 (5), 879–898, 2012. \\

Liu, J., Chen, J.M., Black, T.A., Novak, M.D. E–$\epsilon$ modelling of turbulent air flow downwind of a model forest edge. Boundary-Layer Meteorology 77, 21–44, 1996. \\

Lu, C.H., Fitzjarrald, D.R. Seasonal and diurnal variations of coherent structures over a deciduous forest. Boundary-Layer Meteorology, 69(1–2), 43–69, 1994. \\

Mihalas, D., Weibel-Mihalas, B. Foundations of radiation hydrodynamics. Courier Corporation, 1999. \\

Mirocha, J. D., J. K. Lundquist, and B. Kosović. Implementation of a Nonlinear Subfilter Turbulence Stress Model for Large-Eddy Simulation in the Advanced Research WRF Model. Mon. Wea. Rev., 138, 4212–4228, 2010. \\

Mueller, E., Mell, W., Simeoni, A. Large eddy simulation of forest canopy flow for wildland fire modeling. Canadian Journal of Forest Research 44, 1534–1544, 2014. \\

Patton, E.G., Shaw, R.H., Judd, M.J., Raupach, M.R. Large-eddy simulation of windbreak flow. Boundary-Layer Meteorology 87, 275–307, 1998. \\

Pimont, F., Linn, R.R., Dupuy, J.L., Morvan, D. Effects of vegetation description parameters on forest fire behavior with FIRETEC. Forest Ecology and Management 234(Suppl. 1), S120, 2006. \\

Pimont, F., Dupuy, J.L., Linn, R.R., Dupont, S. Validation of FIRETEC wind-flows over a canopy and a fuel-break. International Journal of Wildland Fire, 18(7), 775-790, 2009. \\

Pimont. F., Dupuy, J.L., Linn, R.R. "Wind and Canopies" In \textit{Wildland Fire Dynamics: Fire Effects and Behavior from a Fluid Dynamics Perspective}, ed. Speer, K., Goodrick, S. (Cambridge University Press: 2022), pages 183-208. \\

Pope, S. B. Turbulent Flows. Cambridge University Press, 2000. \\ 

Potter, B.E. Atmospheric interactions with wildland fire behaviour – I. Basic surface interactions, vertical profiles and synoptic structures. International Journal of Wildland Fire, 21, 779-801, 2012. \\

Raupach, M.R., Bradley, E.F., Ghadiri, H. A wind tunnel investigation into aerodynamic effect of forest clearing on the nesting of Abbott’s Booby on Christmas Island. CSIRO Marine and Atmospheric Research, Centre for Environmental Mechanics, Technical Report T12 1987. (Canberra, Australia) \\

Reisner, JM, Wynne S, Margolin L, Linn RR. Coupled atmospheric– fire modeling employing the method of averages. Monthly Weather Review 128, 3683–3691, 2000a.  \\

Reisner, JM, Knoll DA, Mousseau VA, Linn RR. New numerical approaches for coupled atmosphere–fire models. In ‘Proceedings of the Third Symposium on Fire and Forest Meteorology’, January 2000, Long Beach, CA. pp. 11–14, 2000b. (American Meteorology Society: Boston, MA). \\

Shaw, R. H., Patton, E. Canopy element influences on resolved- and subgrid- scale energy within a large-eddy simulation. Agricultural And Forest Meteorology, 115, 5-17, 2003. \\

Shaw, R.H., Schumann, U. Large-eddy simulation of turbulent flow above and within a forest. Boundary-Layer Meteorology 61, 47–64, 1992. \\

Shaw, R.H., Den Hartog, G., Neumann, H.H. Influence of foliar density and thermal stability on profiles of Reynolds stress and turbulence intensity in a deciduous forest. Boundary-Layer Meteorology 45, 391–409, 1988. \\

Smagorinsky, J., General circulation experiments with the primitive equations: I. The basic experiment. Monthly weather review, 91(3), 99-164, 1963.\\

Smolarkiewicz, P., Margolin, L. MPDATA: A finite difference solver for geophysical flows. Journal of Computational Physics 104(2), 459-480, 1998. \\

Su, H.B., Shaw, R.H., Paw, U.K.T., Moeng, C.H., Sullivan, P.P. Turbulent statistics of neutrally stratified flow within and above a sparse for- est from large-eddy simulation and field observations. Boundary-Layer Meteorology 88, 363–397, 1998. \\

Su, H.B., Shaw, R.H., Paw, U.K.T. Two-point correlation analysis of neutrally stratified flow within and above a forest from large-eddy simulation. Boundary-Layer Meteorology 94, 423–460, 2000. \\

Sullivan, P., Day, M., Pollard, A. Enhanced VITA technique for turbulent structure identification. Exp. Fluids, 18, 10-16, 1994 \\

Taghinia., J. “Sub-grid scale mmoodeling in large eddy simulation with variable eddy-viscosity
coefficient.” PhD thesis. Aalto University, 2015.

Watanabe, T. Large-eddy simulation of coherent turbulence structures associated with scalar ramps over plant canopies. Boundary-Layer Meteorology 112, 307–341, 2004. \\

Wilczak, J. M. Large-scale eddies in the unstably stratified atmospheric surface layer. Part I: Velocity and temperature structures. Journal of Atmospheric Science, 41, 3537-3550, 1984. \\

Yang, B., Raupach, M., Shaw, R.H., Paw, U.K.T., Morse, A.P. Large-eddy simulation of turbulent flows across a forest edge. Part I: Flow statistics. Boundary-Layer Meteorology 120, 377–412, 2006a. \\

Yang, B., Morse, A.P., Shaw, R.H., Paw, U.K.T. Large-eddy simulation of turbulent flow across a forest edge. Part II: Momentum and turbulence kinetic energy budgets. Boundary-Layer Meteorology 121, 433–457, 2006b. 


\end{document}